\documentclass[11pt]{iopart}

\usepackage{graphicx}  
\usepackage{placeins}
\usepackage{lineno}
\usepackage{hhline,multirow,float}
\expandafter\let\csname equation*\endcsname\relax
\expandafter\let\csname endequation*\endcsname\relax
\usepackage{amsmath}
\usepackage{cite} 
\newcommand{\mgo}{Mg$_{0.97}$Co$_{0.03}$O}
\newcommand{\hh}{\hspace*{5.00mm}}

\begin{document}
	
%	\linenumbers
	
%\letter{Confinement of Magnetic Monopoles in Quantum Spin Ice}
%\title[Confinement of Magnetic Monopoles in Quantum Spin Ice]{Confinement of Magnetic Monopoles in Quantum Spin Ice}
\article[P.M.~Sarte et al.]{Article In Memoriam of Professor Roger A. Cowley}{Magnetic Fluctuations and the Spin-Orbit Interaction in Mott Insulating CoO}

\author{P~M~Sarte,$^{1,2,3,4,*}$ S~D~Wilson,$^{1,2}$ J~P~Attfield,$^{3,4}$ \& C~Stock$^{4,5}$}

\address{$^{1}$California NanoSystems Institute, University of California, Santa Barbara, California 93106-6105, USA} 
\address{$^{2}$Materials Department, University of California, Santa Barbara, California 93106-5050, USA} 
\address{$^{3}$School of Chemistry, University of Edinburgh, Edinburgh EH9~3FJ, United Kingdom} 
\address{$^{4}$Centre for Science at Extreme Conditions, University of Edinburgh, Edinburgh EH9~3FD, United Kingdom}
\address{$^{5}$School of Physics and Astronomy, University of Edinburgh, Edinburgh EH9 3FD, United Kingdom}
\ead{\mailto{pmsarte@ucsb.edu}}
%\author{Content \& Services Team}
%\address{IOP Publishing, Temple Circus, Temple Way, Bristol BS1 6HG, UK}
%\ead{submissions@iop.org}
\vspace{10pt}

\begin{abstract}

Motivated by the presence of an unquenched orbital angular momentum in CoO, a team at Chalk River, including a recently hired research officer Roger Cowley, performed the first inelastic neutron scattering experiments on the classic Mott insulator [Sakurai \emph{et al.} 1968 Phys. Rev. $\mathbf{167}$ 510]. Despite identifying two magnon modes at the zone boundary, the team was unable to parameterise the low energy magnetic excitation spectrum below $T\rm{_{N}}$ using conventional pseudo-bosonic approaches, instead achieving only qualitative agreement. It would not be for another 40 years that Roger, now at Oxford and motivated by the discovery of the high-$T_{c}$ cuprate superconductors [Bednorz \& Muller 1986 Z. Phys. B $\mathbf{64}$ 189], would make another attempt at the parameterisation of the magnetic excitation spectrum that had previously alluded him at the start of his career. Upon his return to CoO, Roger found a system embroiled in controversy, with some of its most fundamental parameters still remaining undetermined. Faced with such a formidable task, Roger performed a series of inelastic neutron scattering experiments in the early 2010's on both CoO and a magnetically dilute structural analogue \mgo. These experiments would prove instrumental in the determination of both single-ion [Cowley \emph{et al.} 2013 Phys. Rev. B $\mathbf{88}$ 205117] and cooperative magnetic parameters [Sarte \emph{et al.} 2018 Phys. Rev. B $\mathbf{98}$ 024415] for CoO. Both these sets of parameters would eventually be used in a spin-orbit exciton model [Sarte \emph{et al.} 2019 Phys. Rev. B $\mathbf{100}$ 075143], developed by his longtime friend and collaborator Bill~Buyers, to successfully parameterise the complex spectrum that both measured at Chalk River almost 50 years prior. The story of CoO is of one that has come full circle, one filled with both spectacular failures and intermittent, yet profound, little victories.  
\end{abstract}
%\noindent{\it Keywords}: magnetic materials, inelastic neutron scattering, magnetic properties, magnetic monopoles, spin ice
\maketitle
\ioptwocol
\onecolumn
\section{Introduction}~\label{sec:introduction} 
\begin{figure}[htb!]
	\centering
	\scalebox{0.22}{\includegraphics{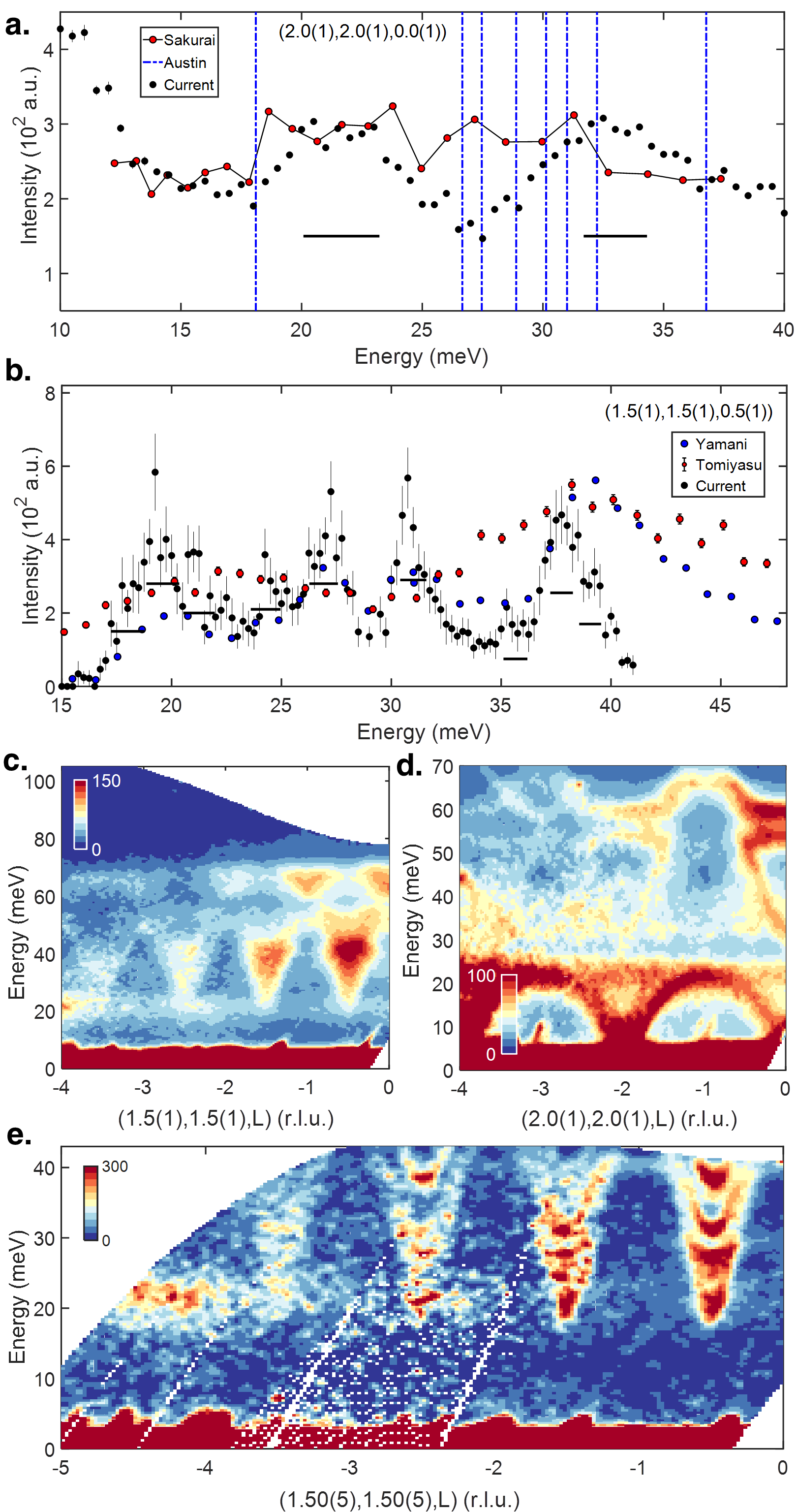}}
	\caption{Comparison of $\mathbf{Q}$-integrated cuts of inelastic neutron scattering data on single crystal CoO reported in the literature~\cite{Sakurai167:68,tomiyasu06:75,yamani403:08,yamani88:10} and recent scattering measurements performed on MERLIN~\cite{BEWLEY06:385} at 5~K by Sarte \emph{et al.}~\cite{Sarte19:100} with an $E_{i}$ = 75~meV and 45~meV at the magnetic zone (a) boundary and (b) centre, respectively.  Solid lines in (a) indicate the location of excitations previously determined by IR spectroscopy~\cite{Austin70:33}. Horizontal bars indicate instrumental resolution~\cite{pychop}. $(\mathbf{Q},E)$ slices of CoO measured on MERLIN at 5~K with an $E_{i}$ of (c) 110 meV, (d) 75 meV, and (e) 45 meV reported by Sarte \emph{et al.}~\cite{Sarte19:100} illustrates an excitation spectrum that is highly structured in both energy and momentum. All $(\mathbf{Q},E)$ slices have been folded along [001]. Adapted from Reference~\cite{Sarte19:100} with permission.}  
	\label{fig:inelastic}
\end{figure}

\begin{figure}[htb!]
	\centering
	\scalebox{0.55}{\includegraphics{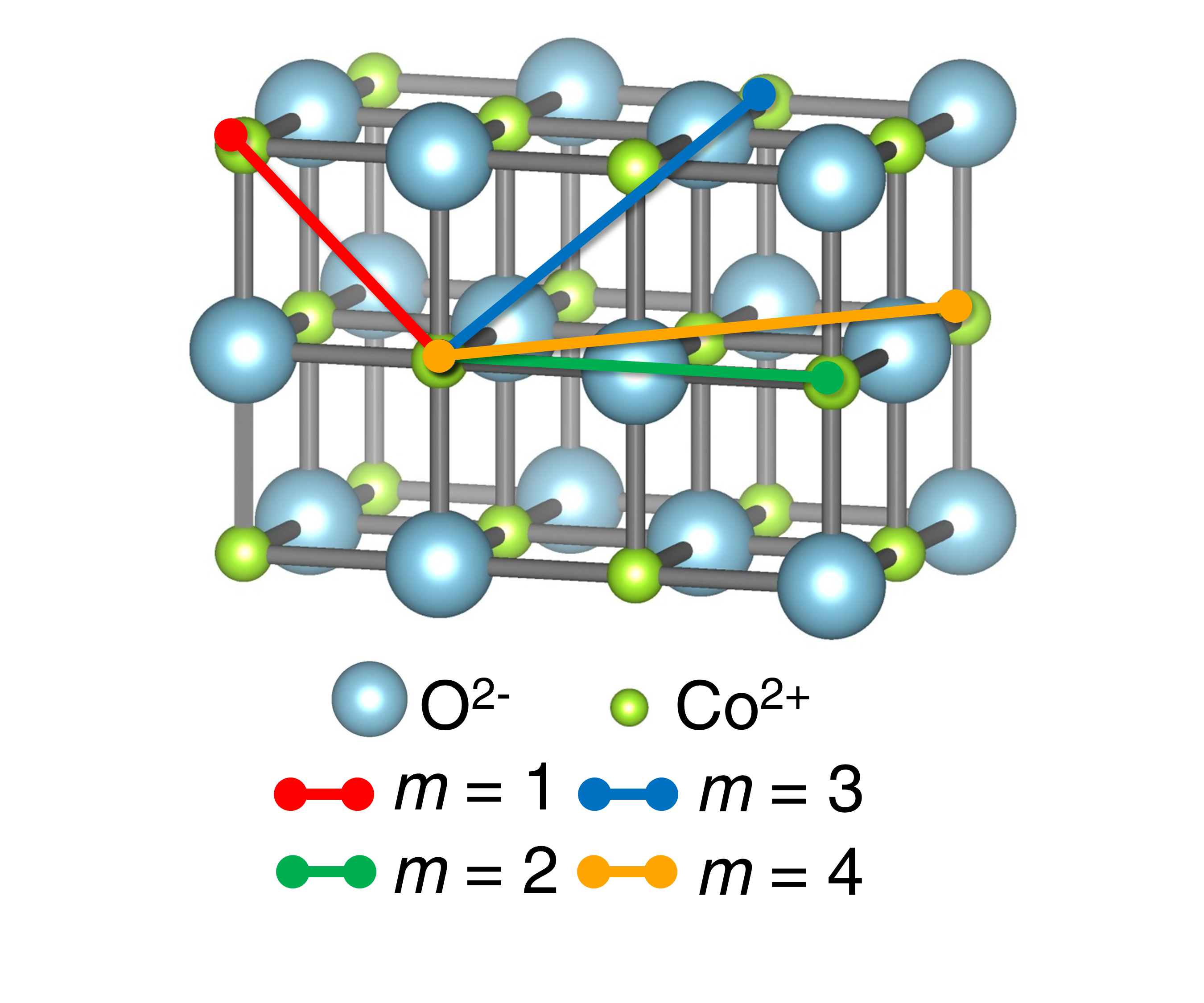}}
	\caption{Isometric view of the first four coordination shells of the high temperature CoO rocksalt ($Fm\bar{3}m$) structure~\cite{Sakurai167:68,jauch01:64,tombs50:165,klemm33:210,Roth58:110,deng10:96}. Adapted from Reference~\cite{Sarte19:100} with permission.}  
	\label{fig:rocksalt}
\end{figure}

\hh In the years following Brockhouse's invention of the neutron triple axis spectrometer~\cite{Shirane:book,svensson04:15,cowley05:51}, Chalk River was considered a premier neutron scattering facility, attracting some of the most brilliant scientific minds of the 20$\rm{^{th}}$ century~\cite{root03:14,powell90:1,banks18:29,greedan18:74}. Many of whom would go on to have long and distinguished careers, whose profound influences are still felt throughout modern physics today~\cite{banks19:4}. One such scientist was Roger~A.~Cowley, whose 6 year tenure (1964-1970) at Chalk River would prove to be particularly influential on the newly minted PhD. A period in his career that would see both high productivity and the establishment of lifelong collaborations. One particularly influential collaborator would be fellow Brit  W.~J.~L.~(Bill)~Buyers, with whom Roger would study magnetic excitations in a wide variety of materials~\cite{birgeneau17:63}. \\
\hh One such material was the classic Mott insulator CoO. First synthesised by Klemm and Sch{\"u}th~\cite{klemm33:210}, the $3d$ metal monoxide garnered significant attention during Roger's tenure at Chalk River due to a combination of its long range antiferromagnetic order~\cite{Roth58:110,vanLaar65:138,vanLaar66:141,trombe51:12} and in particular its unquenched orbital angular momentum~\cite{Abragam:book,kanamori17:57,kanamori17:57_2,Uchida19:64}. CoO has been of interest, more recently, in the context of exchange bias~\cite{Nogues99:192,Berkowitz99:200} in thin films showing a large dependence of properties with thickness~\cite{Zaag84:00}.  Motivated by the unknown role that this unquenched momentum played on the monoxide's magnetic properties including anisotropy, Roger and Bill, along with J.~Sakurai and G.~Dolling, performed the first inelastic neutron scattering experiments on CoO~\cite{Sakurai167:68} . Despite identifying two magnon modes at the magnetic zone boundary (Fig.~\ref{fig:inelastic}(a)), the team at Chalk River was unable to parameterise the magnetic excitation spectrum below $T\rm{_{N}}$ using linear spin wave theory based on the conventional pseudo-bosonic (Holstein-Primakoff) approach~\cite{holstein40:58,Grover65:140,bozorth60:118,cooper69:40}, instead achieving only qualitative agreement. As will be a recurring theme for CoO, this failure in the parameterisation of its magnetic excitations was simply symptomatic of the significant complexity underlying the magnetism in a seemingly ``simple'' monoxide~\cite{feygenson11:83,Wagner75:2,Archer08:78,Cox:book,tomiyasu06:75,yamani403:08,yamani88:10,daniel69:177,lines65:137,yamada74:36}. \\
\hh During the next few decades following the Chalk River experiment, the enigma that is CoO would slowly be revealed. Possessing a simple rock-salt $Fm\bar{3}m$ structure (Fig.~\ref{fig:rocksalt}) at room temperature with a primitive unit cell containing 15 valence electrons~\cite{Sakurai167:68,jauch01:64,tombs50:165}, conventional band theory would predict metallic behaviour~\cite{mattheiss72:5,hugel86:138,norman89:40}. Instead, due to the presence of strong electronic correlations~\cite{terakura84:30}, CoO is a good insulator ($\rho\sim10^{8}~\Omega\cdot$cm at room temperature~\cite{Austin67:90,fisher66:44}) with a band gap of $\sim$2.5~eV~\cite{vanElp44:91}, with evidence for metallic behaviour being found only under extremely high pressures~\cite{Cohen97:275}. In contrast to the ferromagnetism predicted by general band coupling models~\cite{deng10:96,walsh08:100,coey05:4,dalpian06:138}, CoO upon cooling below $T\rm{_{N}}=293$~K undergoes a structural distortion as it assumes long range antiferromagnetic order~\cite{Roth58:110,vanLaar65:138,vanLaar66:141,Dalverny10:114,zeeman65:18,Greenwald53:6,ding06:74}, both of whom have proven controversial~\cite{jauch01:64,saito66:21,pickett89:61,germann74:61,Herrmann78:11}, with such controversy extending to the value of the band gap~\cite{vanElp44:91,Pratt116:59,Powell2:70,Orgel23:55,Shen42:90,YOUMBI14:621,Boussendel10:81,Dalverny10:114,Gillen13:25}, the classification of the antiferromagnetic transition~\cite{olega09:80,mukramel76:13,Bak76:36,silinsky81:24,salamon70:2,srinivasan83:28,lee16:93}, and the identity of the universality class~\cite{salamon70:2,massot08:77,caerels96:6,rechtin70:24}. \\
\hh Ultimately, the greatest source of controversy concerned the paramaterisation of the low energy magnetic fluctuations as determined by the value for both the splittings of the single-ion levels and their interaction parameters~\cite{tomiyasu06:75,feygenson11:83,daniel69:177,chou76:13,El_Batanouny_2002,satoh17:8,keshavarz18:97,rechtin72:5,Smart:book}, despite the formalism to describe both having been already well-established by Kanamori~\cite{kanamori17:57,kanamori17:57_2} and Tachiki~\cite{tachiki64:19} years before the Chalk River experiment. In the case of the crystal field scheme and its splitting parameter $\Delta$, difficulties arose due to limitations inherent to optical techniques~\cite{vanElp44:91,Pratt116:59,degraff237:98,Lorenzana95:74,saitoh01:410,gruninger02:418,liehr57:106}. As will be described later, significant entanglement of the spin-orbit split manifolds due to the presence of multiple far-reaching large exchange interactions with values comparable to that of spin-orbit coupling prevented the extraction of both the spin-orbit coupling constant $\lambda$ and exchange constants $J$ using conventional approaches~\cite{tomiyasu06:75,yamani88:10,yamani403:08,Sakurai167:68,Sarte98:18,kanamori17:57,kanamori17:57_2}.  In the face of a seemingly Herculean task, little progress was seen for almost another 20 years~\cite{feygenson11:83}. \\
\hh It wouldn't be until 1986 with the discovery of high temperature superconductivity in the cuprates by Bednorz and Mueller~\cite{Bednorz64:86} that there would be rejuvenated interest from both Roger and the community as a whole in their parent compounds, the Mott insulators~\cite{Shen42:90,lee06:78,vanElp44:91,brandow76:10,pickett89:61,hufner94:43}. While interest in these Mott insulators has been sustained by the discovery of a wealth of exotic physical phenomena~\cite{Phillips06:321,roy:book,ming17:119,cao16:7,Kim09:1323,jackeli09:102,wang11:106,seo19:122} ranging from metal-to-insulator transitions to quantum criticality to colossal magneto-resistance, the $3d$ transition metal monoxide Mott insulators, including CoO, have garnered particular interest due to their structural ``simplicity'', being often used as toy models where theorists could apply their newest approaches~\cite{Wdowik07:75,savrasov03:90,dudarev00:61,massidda99:82,anisimov91:44,terakura84:30,dufek94:49,Parida18:123}.\\
\hh In a sense, Roger's renewed interest in CoO was ultimately driven by the same motivation. He believed that CoO,  with its unquenched orbital angular momentum, could be used as a tool in the establishment of a complete and comprehensive theoretical framework to describe the electronic structure of the cuprates and ultimately their low temperature properties~\cite{stock05:71,stock07:75,stock10:82,Cowley88:13}. Troubled by the ever increasing complexity of the Hamiltonians employed in conventional linear spin wave theory approaches to describe such systems~\cite{wang13:3,RAPP01:324,buyers06:62,coldea01:86,roger89:39,macdonald88:37,wysin15:book}, Roger desired a framework based on a minimalist Hamiltonian given by 
\begin{equation}
\hat{\mathcal{H}} = \sum\limits_{i}\hat{\mathcal{H}}{\rm{_{CF}}}(i) + \sum\limits_{ij}J_{ij}\hat{\mathbf{S}}_{i} \cdot \hat{\mathbf{S}}_{j}, 
\label{eq:hamiltonian}
\end{equation}
\noindent describing explicitly both single-ion and cooperative magnetism \emph{via} Heisenberg exchange. Despite recent inelastic neutron scattering experiments reported by Tomiyasu \& Itoh~\cite{tomiyasu06:75} and Yamani \emph{et al.}~\cite{yamani403:08,yamani88:10} revealing a magnetic excitation spectrum that was much more complex than what was first measured by the team at Chalk River (Fig.~\ref{fig:inelastic}(b)) --- to be later confirmed by his own measurements~\cite{Sarte19:100} (Figs.~\ref{fig:inelastic}(c)-(f)) --- Roger still believed it was possible that the low energy magnetic fluctuations of a system that had alluded him decades prior could be fully parameterised utilising such a minimalist approach. From Roger's perspective, the use of such a Hamiltonian would provide a much clearer understanding of the complex low temperature magnetism in systems such as CoO by permitting a direct input of spin-orbit, anisotropy, and magnetic exchange parameters, instead of relying on indirect contributions based on expressions derived from higher order perturbation theory~\cite{datta12:85,xiao13:87,khomskii:book,Arts:book}. 

%~\cite{vanElp44:91}
%~\cite{degraff237:98} 
%~\cite{Pratt116:59}
%~\cite{Shen42:90} 
%~\cite{Powell2:70} 
%~\cite{Orgel23:55} 
%~\cite{Sakurai167:68} 
%~\cite{Uchida19:64}

\section{Single-Ion Physics of CoO}

\hh Despite a myriad of both experimental and theoretical studies spanning decades and a rejuvenated interest in CoO spurred on by the cuprates, Roger was still faced with the fact that both terms comprising $\hat{\mathcal{H}}$  in Eq.~\ref{eq:hamiltonian} were still not well-understood at the turn of the new millennium~\cite{tomiyasu06:75,jauch01:64,degraff237:98,feygenson11:83,Wdowik07:75,Farztdinov65:7,Herrmann78:11}. It was at this time that he realised that in order for his vision to come to fruition, the single-ion physics of the $3d^{7}$ Co$^{2+}$ in CoO would need to be first thoroughly understood and ultimately parameterised, before addressing the additional complications arising from cooperative magnetism~\cite{tomiyasu06:75,feygenson11:83,yamani403:08,yamani88:10,datta12:85,alben69:40,Austin70:3,alben69:184}. Such a re-investigation of the single-ion physics would be accomplished experimentally with the technique of neutron scattering~\cite{Cowley88:13}. In contrast to optical spectroscopy corresponding to the technique of choice at the time, neutron scattering is not limited by the selection rules of $\Delta l = \pm 1$~\cite{Lorenzana95:74,Loudon:book}, reducing the difficulty in measuring the often subtle $dd$ transitions ($\Delta l = 0$)~\cite{saitoh01:410,gruninger02:418}. \\
\hh In this section, we describe how Cowley \emph{et al.}~\cite{Cowley88:13} employed neutron scattering to parameterise the two most dominant contributions to $\hat{\mathcal{H}}_{CF}$, thereby laying the foundation for the significant progress that would be seen throughout the next decade in the understanding the low energy magnetic excitations in CoO.       \\
\hh In the form presented in Eq.~\ref{eq:hamiltonian}, $\hat{\mathcal{H}}$ is valid only in the paramagnetic regime~\cite{holden71:4,buyers71:4}. In the case of $T<T\rm{_{N}}$, it can be shown~\cite{buyers75:11,holden74:9} that the introduction of a contribution $\hat{\mathcal{H}}_{MF}$ from the mean molecular field stemming from long range magnetic order that is given by 
\begin{equation}
\hat{\mathcal{H}}_{MF}=\sum_{i}H_{MF}(i)\hat{S}_{z}, 
\label{eq:MF} 
\end{equation}
\noindent where $H_{MF}(i)$ is the molecular field constant for site $i$ that is defined as
\begin{equation}
H_{MF}(i) = 2\sum\limits_{i>j}J(ij)\langle \hat{S}_{z}(j)\rangle,
\label{eq:molecular_field_2} 
\end{equation}
\noindent to Eq.~\ref{eq:hamiltonian} allows $\hat{\mathcal{H}}$ to be rewritten as a sum of a single-ion ($\hat{\mathcal{H}}_{1}$) and an inter-ion ($\hat{\mathcal{H}}_{2}$) term given by
\begin{equation}
\hat{\mathcal{H}}_{1} = \sum\limits_{i}\hat{\mathcal{H}}_{CF}(i) + \sum\limits_{i}\hat{S}_{z}(i)\left(2\sum\limits_{i>j}J(ij)\langle \hat{S}_{z}(j)\rangle\right),
\label{single_ion}
\end{equation}
\noindent and
\begin{equation}
\hat{\mathcal{H}}_{2} = \sum\limits_{ij}J(ij)\hat{S}_{z}(i)[\hat{S}_{z}(j)-2\langle \hat{S}_{z}(j)\rangle] + \frac{1}{2}\sum\limits_{ij}J(ij)[\hat{S}_{+}(i)\hat{S}_{-}(j) + \hat{S}_{-}(i)\hat{S}_{+}(j)],
\label{Inter_ion}
\end{equation}
\noindent respectively, where $\langle \hat{S}_{z}(j)\rangle$ denotes a thermal average given by $\langle \hat{S}_{\eta} \rangle = \sum\limits_{n}f_{n}\langle n|\hat{S}_{\eta}|n\rangle$,
weighted by the Boltzmann thermal population factor  $f_{n}$. The importance of this particular separation of $\hat{\mathcal{H}}$ into two terms will later become apparent when discussing the spin-orbit exciton model~\cite{buyers75:11,Sarte19:100}.  The expression for the single-ion Hamiltonian of CoO can then be rewritten as 
\begin{equation}
\hat{\mathcal{H}}{\rm{_{1}}} = \hat{\mathcal{H}}_{CF}+ \hat{\mathcal{H}}_{MF}=
(\hat{\mathcal{H}}_{CEF}+\hat{\mathcal{H}}_{SO} + \hat{\mathcal{H}}_{dis})+ \hat{\mathcal{H}}_{MF},
\label{eq:SI}
\end{equation}
 \noindent consisting of the individual contributions from the crystalline electric field, spin-orbit coupling, structural distortion, and mean molecular field, respectively~\cite{Sakurai167:68,Cowley88:13,Sarte98:18,Sarte19:100,tomiyasu06:75,buyers71:4,holden71:4}.\\
\subsection{Crystalline electric field, Tanabe-Sugano \& determination of $10Dq$}
\begin{figure}[htb!]
	\centering
	\scalebox{0.55}{\includegraphics{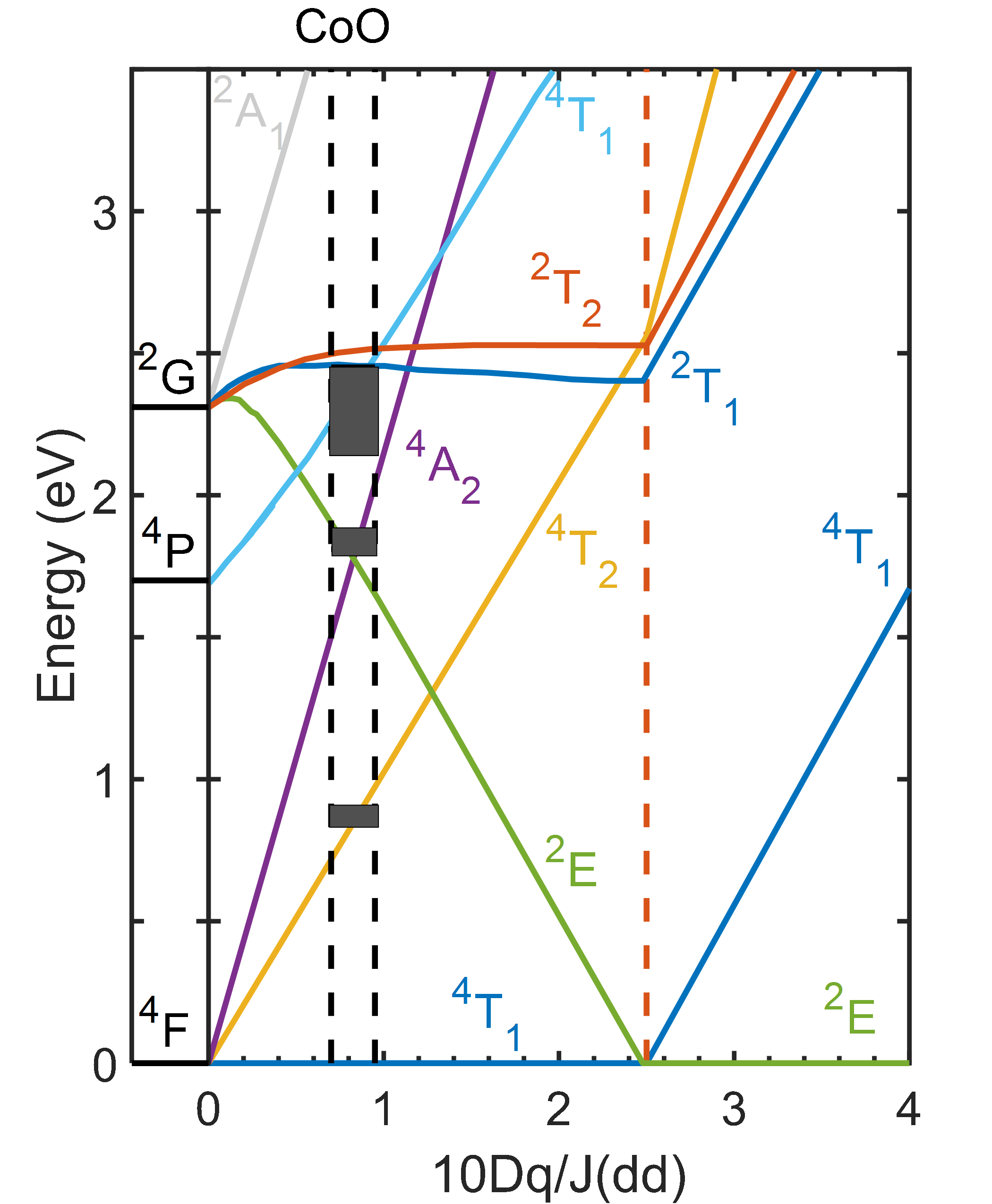}}
	\caption{Tanabe-Sugano diagram~\cite{Tanabe54:9,Tanabe54:9_2} for $d^{7}$ Co$^{2+}$ in octahedral coordination calculated by Cowley \emph{et al.}~\cite{Cowley88:13}. Shaded rectangles correspond to excitations for cubic CoO experimentally measured on MAPS~\cite{ewings19:90} at room temperature, with the height and width corresponding to experimental error in energy and the statistical error of the refined value for $10Dq/J(dd)$, respectively. The red dashed line at $10Dq/J(dd)\sim2.5$ denotes the spin crossover from (left) a high-spin $S=\frac{3}{2}$, $^{4}T_{1}$ orbital configuration to (right) the low-spin $S=\frac{1}{2}$, $^{2}E$. Adapted from References~\cite{Cowley88:13,Sarte19:100} with permission.}  
	\label{fig:CEF}
\end{figure}

\hh At the time of Roger's neutron scattering experiments, the crystal field scheme and the formalism to address $\hat{\mathcal{H}}_{CEF}$ in CoO was still controversial, stemming from the limitations of optical measurements and from the question concerning which theoretical framework was appropriate, respectively~\cite{degraff237:98,Pratt116:59,Kemp89:1,gorschluter94:49,hassel95:240,johansen77:33,vanElp44:91,schlapp32:42,kambe52:7}. At one extreme was \emph{weak crystal field theory}, where the crystal field splitting of the free-ion states was much smaller than the energy difference between these free-ion states, while at the other extreme was \emph{strong crystal field theory}, where the opposite was true~\cite{Abragam:book}.  For the former case,  $\hat{\mathcal{H}}_{CEF}$ is treated as a perturbation to the free-ion states, with a complete set of commuting observables of: $\hat{L}^{2}$, $\hat{L}_{z}$, $\hat{S}^{2}$ and $\hat{S}_{z}$ with corresponding good quantum numbers $L$, $m_{L}$, $s$ and $m_{s}$. For the case of octahedral coordination~\cite{hutchings64:16,WALTER84:45}, $\hat{\mathcal{H}}_{CEF}$ can be expressed in terms of the Stevens operators $\hat{\mathcal{O}}^{0}_{4}$ and $\hat{\mathcal{O}}^{4}_{4}$, and the numerical coefficient $B_{4}<0$ as 
\begin{equation}
\hat{\mathcal{H}}_{CEF} = B_{4}\left(\hat{\mathcal{O}}^{0}_{4} + 5\hat{\mathcal{O}}^{4}_{4}\right). 
\label{eq:CF}
\end{equation}     
\noindent In contrast, the basis in \emph{strong crystal field theory} corresponds to the $|t_{2g}\rangle$ and $|e_{g}\rangle$ orbitals. In the case of CoO, whereas in \emph{weak crystal field theory}, where Hund's rules are applied to the degenerate $3d$ orbitals to determine the $^{4}F$ ($L = 3$, $s=\frac{3}{2}$) ground state free-ion electronic configuration, in the case of \emph{strong crystal field theory}, Hund's rules are applied to the split $3d$ orbitals to determine the ground state orbital electronic configuration $^{2}E$ ($s=\frac{1}{2}$). \\
\hh Armed with high quality single crystals and a high flux of epithermal neutrons present on the direct geometry chopper spectrometer MAPS at the ISIS spallation source~\cite{ewings19:90,EWINGS16:834}, three distinct $dd$ transitions at  $0.870(9)$~eV, $1.84(3)$~eV, and $2.30(15)$~eV were identified at 300~K~\cite{Cowley88:13}. Their identity being confirmed by both eliminating a non-electronic background component, and by noting the $Q$ and temperature dependence of the excitations of interest. \\
\hh Having experimentally determined the $dd$ transitions, the parameterisation of $\hat{\mathcal{H}}_{CEF}$ was accomplished by a comparison of these transitions to the energy levels of Co$^{2+}$ calculated from electrostatic matrices and character tables based on the formalism previously established by Tanabe and Sugano~\cite{Tanabe54:9,Tanabe54:9_2}. These electrostatic matrices are defined by both the crystal field splitting $10~Dq$ and electron-electron interaction parameters given by $B$, $C$ or $J(dd)$, $C(dd)$ in Racah or Hubbard formalism, respectively~\cite{Abragam:book,griffith:book,MCCLURE59:9}. \\
\hh As illustrated in Fig.~\ref{fig:CEF}, the experimental data was successfully reproduced~\cite{Cowley88:13} by the Tanabe-Sugano diagram with values for $J(dd)$, $C(dd)$, and $10Dq$ of $1.3(2)$~eV, $0.49(10)$~eV, and $0.94(10)$~eV, respectively, with both the values of $J(dd)$ and its ratio to $C(dd)$ showing excellent agreement with previously calculated values~\cite{MCCLURE59:9,dermarel88:37,anisimov91:44}.  The ratio $\frac{10Dq}{J(dd)}$ = 0.72(18) confirmed that Co$^{2+}$ in CoO lies significantly below the spin-crossover value of $\frac{10Dq}{J(dd)}\sim$2.5, assuming a high spin ($s=\frac{3}{2}$) orbital triplet $^{4}T_{1}$ ground state, despite a value of $\sim$1~eV for $10Dq$, placing CoO in the so-called \emph{weak field-intermediate crystal field regime}~\cite{Abragam:book}. In this particular regime, the complete set of commuting observables is identical to the set found in \emph{weak crystal field theory}. In the case of magnetic properties, the values of $L$ and $s$ are fixed to those in the $^{4}F$ Co$^{2+}$ free-ion ground state  ($L=3,s=\frac{3}{2}$).  Diagonalisation of $\hat{\mathcal{H}}_{CEF}$ for the $^{4}F$ free-ion ground state in an octahedral coordination (Eq.~\ref{eq:CF}) yields an orbital triplet ground state $^{4}T_{1}$, an excited orbital triplet $^{4}T_{2}$, and an orbital singlet $^{4}A_{2}$, where $\Delta(^{4}T_{1}\rightarrow$$^{4}T_{2})=480B_{4}$ and  $\Delta(^{4}T_{2}\rightarrow$ $^{4}A_{2})=600B_{4}$, where the Stevens factor $B_{4}$ is related to the crystal field splitting by $10Dq=400 B_{4}$~\cite{Abragam:book,Sarte98:18_2,WALTER84:45,wallington15:92,hutchings64:16}.\\
\hh The identification of the 1.84(3)~eV excitation as the $^{4}T_{1}\rightarrow$$^{4}A_{2}$ transition  was particularly noteworthy because there were two possibilities for the second excited state, corresponding to either $^{4}A_{2}$ or $^{2}E$~\cite{degraff237:98,vanElp44:91}. Both transitions are dipolar forbidden with both exhibiting transition matrix elements $|\langle|0|\hat{\mathbf{M}}_{\pm,z}|f\rangle| = 0$, where $\hat{\mathbf{M}}_{\pm,z} = \hat{\mathbf{L}}_{\pm,z} + 2\hat{\mathbf{S}}_{\pm,z}$~\cite{ross17:95,Shirane:book}, and thus could not distinguished based on intensities alone in the dipolar approximation~\cite{sakurai:book}. Instead, Cowley \emph{et al.}~\cite{Cowley88:13} noted that the $^{4}T_{1}\rightarrow$$^{4}A_{2}$ transition corresponded to a large quadrupolar matrix element which had been previously predicted, but at the time had not been resolved experimentally~\cite{haverkort07:99,larson07:99}. 

\subsection{Spin-orbit coupling}

\hh The low value of $Z$ for Co$^{2+}$ places CoO in the Russell-Saunders $L$-$S$ coupling scheme where spin-orbit coupling is treated as a pertubration of the form $\hat{\mathcal{H}}_{SO}=\lambda \hat{\mathbf{L}} \cdot \hat{\mathbf{S}}$ to the $|L=3,m_{L},s=\frac{3}{2},m_{s}\rangle$ basis defined by $\hat{\mathcal{H}}_{CEF}$ in the low energy limit~\cite{kanamori17:57,kanamori17:57_2,tomiyasu06:75,wallington15:92,buyers71:4,holden71:4,low58:109}. As was the case for $\hat{\mathcal{H}}_{CEF}$, where the $^{4}F$ free-ion state was exclusively considered, a similar approach may be taken for $\hat{\mathcal{H}}_{SO}$~\cite{schlapp32:42,kambe52:7}. Since $10Dq\sim$1~eV~\cite{Cowley88:13}, the magnetic properties must solely be determined by the $^{4}T_{1}$ ground state. Such a restriction of the analysis to the $^{4}T_{1}$ ground state manifold requires a projection from the original $|L=3,m_{L}\rangle$ basis onto a smaller basis $|l=1,m_{l}\rangle$, yielding a new projected spin-orbit Hamiltonian~\cite{buyers71:4,abragam51:205,abragam51:206} given by 
\begin{equation}
\hat{\mathcal{H}}_{SO} = \alpha\lambda\hat{\mathbf{l}}\cdot\hat{\mathbf{S}}, 
\label{eq:SO_2}
\end{equation}   
\noindent consisting of new orbital angular momentum operators that act on the projected $|l=1,m_{l},s=\frac{3}{2},m_{s}\rangle$ basis, accompanied by a projection factor $\alpha$. Although it was the convention at the time to determine the value of $\alpha$ using representation theory~\cite{Abragam:book}, Sarte \emph{et al.}~\cite{Sarte98:18_2,Jana99:19},  inspired by work done on 4$d$
and 5$d$ transition metal oxides by Stamokostas and Fiete~\cite{stamokostas18:97}, later developed an alternate method based on the matrix representation of angular momentum operators for obtaining these projection factors $\alpha$. \\
\hh Having projected $\hat{\mathbf{L}}$ onto a fictitious operator $\hat{\mathbf{l}}$ with $l =1$, the basis for $\hat{\mathcal{H}}_{SO}$ (Eq.~\ref{eq:SO_2}) is now the 12 $|l=1, m_{l}; s={3\over 2}, m_{s} \rangle$ states. Based on both the Land\'{e} interval rule and the addition theorem~\cite{sakurai:book}, $\hat{\mathcal{H}}_{SO}$ yields three unique effective total angular momentum $\hat{\mathbf{j}} = \hat{\mathbf{l}} + \hat{\mathbf{S}}$ manifolds corresponding to $j\rm{_{eff}}$=${1\over 2}$, ${3 \over 2}$, and ${5 \over 2}$, with their energy eigenvalues given by~\cite{Cowley88:13,wallington15:92,khomskii:book_2,lines63:131,lines65:137} 
\begin{equation}
E = \frac{\alpha\lambda}{2}\left[j{\rm{_{eff}}}(j{\rm{_{eff}}}+1)-s(s+1)-l(l+1)\right].
\end{equation}   
%\noindent In the low temperature limit with $l$ and $s$ being fixed to 1 and $\frac{3}{2}$, respectively, the energy difference between the ground state $j\rm{_{eff}}$=${1\over 2}$ doublet and first excited $j\rm{_{eff}}$=${3\over 2}$ quartet is given by $\frac{3\alpha\lambda}{2}$, whilst the energy difference between the two excited $j\rm{_{eff}}$=${3\over 2}$ and $j\rm{_{eff}}$=${5\over 2}$ manifolds is given by $\frac{5\alpha\lambda}{2}$. \\
\hh Since the energy differences between $j\rm{_{eff}}$ manifolds are fixed, $\lambda$ (in theory) can be determined experimentally by measuring the energy transfer associated with spin-orbit transitions. To accomplish such task, Cowley \emph{et al.}~\cite{Cowley88:13} performed neutron scattering experiments at low temperatures. The choice of neutron scattering limited the measurement of spin-orbit transitions to those transitions associated with $\Delta j\rm{_{eff}} =0, \pm1$~\cite{Shirane:book}; while the use of low temperatures limited the measurement of spin-orbit transitions exclusively out of the ground state. \\

\subsection{Tetragonal distortion \& mean molecular field}

\begin{figure}[htb!]
	\centering
	\scalebox{0.285}{\includegraphics{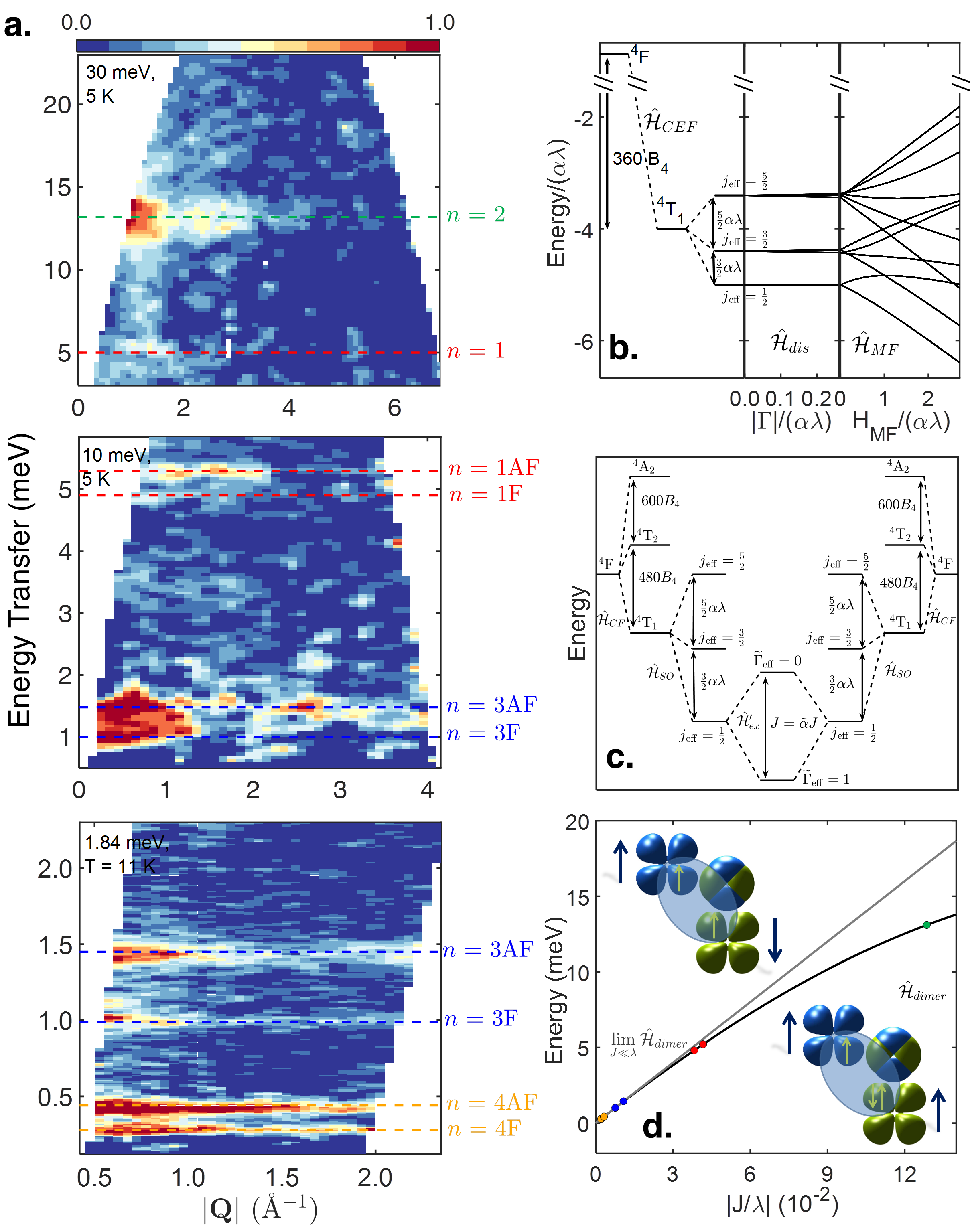}}
	\caption{(a) Background (nonmagnetic MgO) subtracted $S(Q,E)$ maps of Mg$_{0.97}$Co$_{0.03}$O measured on (top) MARI at 5~K with an $E_{i}$=30~meV, (middle) MARI at 5~K with an $E_{i}$=10~meV, and (bottom) IRIS at 11~K with an $E_{f}$ of 1.84~meV revealing seven
low-energy dispersionless magnetic excitations~\cite{Sarte98:18}. (b) Calculated energy eigenvalues as a function of the tetragonal distortion ($\hat{\mathcal{H}}_{dis}$) and the magnetic molecular field ($\hat{\mathcal{H}}_{MF}$) perturbations to the $j\rm{_{eff}}$ manifolds of the $^{4}T_{1}$ ground state for $d^{7}$ Co$^{2+}$ in octahedral coordination. Both the energy eigenvalues and individual parameters are presented to scale. (c) Relevant energy scales for the effective pair Hamiltonian $\hat{\mathcal{H}}_{pair}$ (Eq.~\ref{eq:pair})~\cite{buyers71:4}. (d) Energy splitting of the $j\rm{_{eff}}=\frac{1}{2}$ manifold calculated from the diagonalisation of $\hat{\mathcal{H}}_{pair}$ (black line). The non-linearity constrasts the behaviour predicted by the projection theorem~\cite{Suhonen2007,sakurai:book,Abragam:book} (gray line). (inset) The mechanism for antiferromagnetism (top) and weaker ferromagnetism (bottom) is a result of a combination of the 90$^{\rm{o}}$ Co$^{2+}$$-$O$^{2-}$$-$Co$^{2+}$ exchange pathway and the orbital degree of freedom in the $t_{2g}$ channel on each Co$^{2+}$, in agreement with the predictions of the Goodenough-Kanamori-Anderson rules~\cite{goodenough55:100,goodenough58:6,anderson50:79,kanamori59:10,weihe97:36}. Yellow arrows denote local $t_{2g}$ spin configurations and teal arrows denote total spin configurations on each Co$^{2+}$. Adapted from References~\cite{Sarte98:18,Sarte19:100} with permission.}  
	\label{fig:MARI}
\end{figure}

\hh As described so far, the experimental determination of $\lambda$ appears deceptively simple. Such simplicity is based on two assumptions. The first is that the degeneracy for each $j\rm{_{eff}}$ manifold is maintained, thus restricting $\Delta E\neq 0$ transitions to those with $\Delta j\rm{_{eff}}=\pm1$. The second assumption is that the $j\rm{_{eff}}$ manifolds are not shifted significantly away from the eigenvalues of $\hat{\mathcal{H}}_{SO}$, such that $\Delta E$ remain proportional to $\lambda$. In the case of pure CoO, both assumptions are invalid~\cite{kanamori17:57,kanamori17:57_2,Sakurai167:68,tomiyasu06:75}. \\
\hh As summarised by Fig.~\ref{fig:MARI}(b), the invalidity stems from the contributions from the two remaining terms in $\hat{\mathcal{H}}{\rm{_{1}}}$ that have been conveniently absent in our discussion so far. The first is the distortion Hamiltonian $\hat{\mathcal{H}}_{dis}$, which corresponds to the structural deformation of the unit cell that accompanies long-range antiferromagnetic order~\cite{Roth58:110,vanLaar65:138,vanLaar66:141,Dalverny10:114,zeeman65:18,Greenwald53:6}.  While the exact symmetry of the low temperature phase has proven to be particularly contentious~\cite{jauch01:64}, we have chosen to consider the case of a tetragonal distortion which we will show captures the essential low energy physics.    We note that our inelastic data is not able to uniquely identify the nuclear and magnetic structures.  Utilising symmetry arguments, the influences of a tetragonal distortion is given by~\cite{WALTER84:45,wallington15:92,Cowley73:6,gladney66:146,martel68:46}
\begin{equation}
\hat{\mathcal{H}}_{dis}=B_{2}\hat{\mathcal{O}}^{0}_{2}=\Gamma  \left(\hat{l}^{2}_{z}-{2 \over 3} \right),
\end{equation}
\noindent where $\Gamma$ corresponds to a distortion parameter, and whose presence in $\hat{\mathcal{H}}{\rm{_{1}}}$ results in the removal of the degeneracy of the spin-orbit manifolds~\cite{Ham62:18,jahn37:161,jahn38:164}. \\
\hh The second additional term is the molecular field Hamiltonian $\hat{\mathcal{H}}_{MF}$. Behaving as a Zeeman-like term, $\hat{\mathcal{H}}_{MF}$ breaks time reversal symmetry, thus removing the degeneracy of the individual $j\rm{_{eff}}$ manifolds~\cite{wallington15:92}.  As a first approximation, by considering exclusively a single dominant next-nearest neighbour $180^{\circ}$ Co$^{2+}$-O$^{2-}$-Co$^{2+}$  superexchange pathway with a magnetic exchange constant $J_{2}$, $\hat{\mathcal{H}}_{MF}$ (Eq.~\ref{eq:MF}) can be simplified to~\cite{buyers71:4,holden71:4,Sakurai167:68} 
\begin{equation}
\hat{\mathcal{H}}_{MF}=2z_{2}J_{2}\langle \hat{S_{z}}\rangle \hat{S}_{z},
\end{equation} 
\noindent where $z_{2}$ denotes the number of next-nearest neighbours. As illustrated in Fig.~\ref{fig:MARI}(b), a strong value of this exchange interaction results in significant mixing of spin-orbit split levels~\cite{kanamori17:57,kanamori17:57_2}. In the case of CoO, such entanglement is further exacerbated by the presence of various long range spin-spin superexchange interactions with exchange constants comparable to $\lambda$~\cite{tomiyasu06:75,yamani88:10,yamani403:08,feygenson11:83}. With the additional complications of both complex magnetic ordering and structural distortions~\cite{vanLaar65:138,vanLaar66:141,jauch01:64,tomiyasu04:70}, it quickly becomes apparent how such a deceptively ``simple'' $3d$ monoxide yields such a complex magnetic excitation spectrum. The limitations imposed on approaches based on conventional linear spin wave theory~\cite{holstein40:58,Grover65:140,bozorth60:118,cooper69:40}, combined with the necessity of a multiparameter spin-orbital Hamiltonian incorporating both exchange and spin-orbit coupling to describe such a spectrum, made both the modelling and understanding of the magnetic excitations in CoO particularly elusive for decades~\cite{Sarte19:100,STRUZHKIN93:168,Wagner75:2,kant08:78,fischer09:80,tomiyasu06:75}. 

\subsection{\mgo~\& determination of $\lambda$}

\hh In order to circumvent the problematic $\hat{\mathcal{H}}_{MF}$ contribution, it is important to note that the significant degree of entanglement present in CoO was not inherent to all Co$^{2+}$ magnets, with some possessing a small $H_{MF}$ due to $|J|$ being significantly smaller than $|\lambda|$~\cite{wallington15:92,buyers71:4,holden71:4,Sarte98:18_2,ross17:95,cabrera14:90,Cowley73:6}. In contrast to CoO, these Co$^{2+}$-magnets exhibited a much weaker degree of mixing, such that the spin-orbit split levels remained well-separated in energy. Although limited by the large value of $J_{2}$ in CoO, Cowley \emph{et al.}~\cite{Cowley88:13} effectively eliminated $H_{MF}$ by the chemical dilution of Co$^{2+}$ by non-magnetic Mg$^{2+}$. The resulting dilute monoxide \mgo, in which the majority of Co$^{2+}$ are isolated from one another, does not assume long range magnetic order down to 5~K. Furthermore, as its unit cell remains cubic upon cooling ($\hat{\mathcal{H}}_{dis} = 0$), $\hat{\mathcal{H}}\rm{_{1}}$ for the case of \mgo~can be reduced to $\hat{\mathcal{H}}_{CEF} + \hat{\mathcal{H}}_{SO}$ (Fig.~\ref{fig:MARI}(c)). \\
\hh Neutron scattering experiments on a polycrystalline sample of \mgo~at 5~K on the direct geometry chopper spectrometer MARI~\cite{ANDERSEN1996472,stock10:81} at ISIS identified one distinct excitation at 37.1(5)~meV~\cite{Cowley88:13}. 
This particular excitation exhibited a $Q$ dependence consistent with the Co$^{2+}$ magnetic form factor, thus confirming its origin as electronic, while its FWHM of 6(1)~meV being significantly broader than the instrumental resolution $\sim 2\% \frac{\Delta E}{E_{i}} \sim 2$~meV~\cite{pychop} is consistent with a distribution of different local environments that would be expected from such a dilute monoxide. \\
\hh Having established that the excitation was indeed the $j{\rm{_{eff}}}=\frac{1}{2} \rightarrow j{\rm{_{eff}}}=\frac{3}{2}$ spin-orbit excitation and given the gap $\Delta E = \frac{-9\lambda}{4}$~\cite{buyers71:4,holden71:4,Cowley73:6,Sarte98:18,Sarte98:18_2}, an energy transfer of 37.1(5)~meV yields a value for $\lambda$ of $-16(3)$~meV. Despite not requiring any detailed knowledge of the $Q$ dependence of the spin-excitations, the inclusion of multiple exchange constants, and a structural distortion contribution, the value of $\lambda$ determined through chemical dilution was in fact consistent with values obtained from nominally pure CoO through indirect and model dependent approaches~\cite{Abragam:book,tomiyasu06:75,tomiyasu04:70,kant08:78,vanschooneveld12:116,satoh17:8,Austin70:33,ferguson63:39,hirakawa60:15,thornley65:284}. 

\section{Extraction of Exchange Constants $J$} 

\hh Having parameterised the two largest contributions to $\hat{\mathcal{H}}\rm{_{1}}$, Roger's attention now shifted towards addressing the cooperative magnetism in CoO. Faced with both the aforementioned limitations imposed on conventional linear spin wave theory approaches and a body of often contradictory literature spanning decades~\cite{feygenson11:83,satoh17:8,chou76:13,kant08:78,daniel69:177,Sakurai167:68,kanamori17:57,kanamori17:57_2,tachiki64:19,tomiyasu06:75,El_Batanouny_2002}, Roger knew that the problem of extracting $J$ would need to be accomplished through an alternative approach.  \\
\hh Inspired by neutron inelastic studies on small magnetic clusters ($N\leq4$)~\cite{Furrer13:85,buyers84:30,GUDEL81:106,guedel79:18}, in particular those magnetically diluted systems based on Mn$^{2+}$~\cite{svensson78:49,Falk84:52,Falk87:35,Falk86:56,BREITLING77:6,furrer11:83}, and his previous work on 2D and 3D transition metal fluorides~\cite{cowley77:39,hagen83:28,COWLEY:book,cowley80:22,birgeneau80:21,birgeneau79:50}, Roger speculated that such an alternative approach would be centred on the dilute monoxide \mgo~\cite{Cowley88:13}. Although the absence of both long range magnetic ordering and the accompanying structural distortion would conserve the degeneracy of the $j\rm{_{eff}}$ manifolds, Roger rationalised that these individual manifolds would be split by the superexchange interaction. \\
\hh Motivated by this alternative, and seemingly more ``direct'' method for extracting $J$~\cite{Furrer13:85,haraldsen05:71}, additional inelastic neutron scattering measurements were performed on polycrystalline \mgo~\cite{Sarte98:18}. As illustrated in Fig.~\ref{fig:MARI}(a), measurements at 5~K displayed a hierarchy of seven dispersionless magnetic excitations up to an energy transfer $\Delta E\sim$15~meV. These excitations  exhibited clear modulation in $Q$ that is characteristic of magnetic clusters~\cite{haraldsen05:71,waldmann03:68,tennant97:78,koo02:41}, thus distinguishing them from single-ion dispersionless crystal-field excitations that were previously observed at higher energy transfers~\cite{Cowley88:13}. \\
\hh It was at this time that Roger was faced with two important questions. The first was what were these magnetic clusters? It was clear that these excitations were not single-ion in origin, but it was not clear if these excitations were from interactions between pairs ($N=2$), triples ($N=3$), $etc.$ The second was how would the $J$ constants be extracted? In contrast to $d^{5}$ Mn$^{2+}$, the magnetism for $d^{7}$ Co$^{2+}$ is arguably made much more complicated by the orbital degree of freedom in the $t_{2g}$ channel, requiring multiple projections of angular momentum operators~\cite{kanamori17:57,kanamori17:57_2,buyers71:4,abragam51:205,abragam51:206,Abragam:book,khomskii:book_2,Sarte98:18_2}. \\
\hh In this section, we describe how Roger's collaborators Sarte \emph{et al.}~\cite{Sarte98:18} exploited the paramagnetic nature \mgo~to disentangle the spin exchange and spin-orbit interactions.  It would take almost three additional years for this disentanglement, combined with the interpretation of the energy hierarchy of excitations as Co$^{2+}$ pairwise interactions, to result in the extraction of 7 exchange constants out to the fourth coordination shell in the rocksalt lattice. 

\subsection{Sample characterisation and dominance of Co$^{2+}$ pairs}

\hh Although the value of 3\% Co$^{2+}$ concentration at first appears nonsensical, especially in the context of an often flux-limited technique such as inelastic neutron scattering~\cite{Shirane:book}, this particularly low concentration actually significantly simplifies the magnetism under consideration by restricting the cooperative magnetism in \mgo~to be dominated by pairwise interactions between isolated Co$^{2+}$ pairs~\cite{Furrer13:85}. Such dominance can be rationalised from probabilistic arguments, where it can be shown that the ratio of numbers of $N$ to $N+1$ spin clusters in Mg$_{1-x}$Co$_{x}$O is given by $\frac{1-x}{x}$; thus, for small values of $x$, the number and hence scattering from Co$^{2+}$ pair excitations far outweigh those from larger clusters, as has been observed with dilute Mn$^{2+}$-based magnets for $x<0.05$~\cite{BREITLING77:6,Falk87:35}. \\
\hh Although such arguments provide an excellent rationale for the dominance of pairwise interactions in \mgo, these probabilistic arguments are ultimately based on the assumptions that the Co$^{2+}$ concentration is actually $x=0.03$ and its distribution is homogenous. In order to confirm the value of $x$, its value
was experimentally determined using both x-ray diffraction and DC magnetometry. \\
\hh The question of chemical homogeneity and the possibility of Co$^{2+}$ clustering was addressed by first noting the absence of a ZFC/FC split in the DC susceptibility. Secondly, the powder-averaged dynamic structure factor $S(Q,E)$ for \mgo~was shown to be exhibit completely diffferent features compared to those present in either CoO and MgO. A third and final argument is that the value of $\lambda$ that was extracted from \mgo~agreed very well with the reported values in the literature~\cite{Cowley88:13,Abragam:book,tomiyasu06:75,kant08:78,vanschooneveld12:116}. An agreement only made possible by the disentanglement of the individual spin-orbit manifolds, a process that would hindered by any type of clustering~\cite{Furrer13:85}. \\
\hh Direct experimental confirmation of the absence of any Co$^{2+}$ clustering was accomplished by two vastly different methods. The first involved performing the same scattering experiments at identical temperatures with slightly different incident energies on a second polycrystalline sample of \mgo~that was synthesised by a solution-based (sol-gel) technique~\cite{danks16:3} \emph{in lieu} of the traditional ceramic approach~\cite{Rao93:1,Rao:book}. The second method consisting of EDX-SEM measurements~\cite{SEM} confirmed the uniform distribution of Co$^{2+}$, Mg$^{2+}$, and O$^{2-}$, with no evidence for significant clustering. The resulting elemental analysis confirmed that the Co$^{2+}$ concentration was within experimental error to the values previously deduced by both DC magnetometry and x-ray diffraction measurements. 

 \subsection{Effective pair spin-orbital Hamiltonian}
 
\hh Having established that the cooperative magnetism and thus the low energy response in \mgo~is dominated by Co$^{2+}$ pairs and their pairwise interactions, the discussion now shifts 
to the neutron scattering response of an isolated pair of magnetic ions and how it can be used to extract both the interaction distance $R$ and exchange constant $J$. \\
\hh As summarised by Fig.~\ref{fig:MARI}(c), the low energy fluctuations in \mgo~correspond exclusively to excitations within the ground state doublet, and thus requiring the projection of the spin operator $\hat{\mathbf{S}}$ onto the $j\rm{_{eff}}=\frac{1}{2}$ manfiold~\cite{buyers71:4,holden71:4,Sarte98:18_2,kanamori17:57,kanamori17:57_2,tomiyasu06:75,ross17:95,furrer11:83}. By defining the projected angular momentum operator $\hat{\mathbf{S}}=\beta\hat{\mathbf{j}}$ with $j=\frac{1}{2}$, the interaction energy between a Co$^{2+}$ spin pair given by $\hat{\mathcal{H}}_{ex} = 2J\hat{\mathbf{S}}_{1}\cdot\hat{\mathbf{S}}_{2}$ can be approximated by 
\begin{equation}
\hat{\mathcal{H}}'_{ex} \sim \tilde{\alpha}J\hat{\mathbf{j}}_{1}\cdot\hat{\mathbf{j}}_{2},
\end{equation}  
\noindent where the effective projection factor $\tilde{\alpha} = 2\beta^{2}$ is a proportionality constant between the magnetic exchange constant $J$ and the gap between the triplet ($\Gamma\rm{_{eff}}=1$) and singlet ($\Gamma\rm{_{eff}}=0$) states being given by $\Delta E = \tilde{\alpha}|J|$~\cite{Furrer13:85,haraldsen05:71}. This value of $\tilde{\alpha}$ can be calculated numerically by diagonalising an effective pair Hamiltonian 
\begin{equation}
\hat{\mathcal{H}}_{pair} = \alpha\lambda\hat{\mathbf{l}}_{1}\cdot\hat{\mathbf{S}}_{1}+ \alpha\lambda\hat{\mathbf{l}}_{2}\cdot\hat{\mathbf{S}}_{2} +  2J\hat{\mathbf{S}}_{1}\cdot\hat{\mathbf{S}}_{2},
\label{eq:pair}
\end{equation}
\noindent describing the superexchange interaction of the spin-orbit split manifolds of a Co$^{2+}$ pair in the non-distorted ($\hat{\mathcal{H}}_{dis} = 0$) and magnetically dilute paramagnetic ($\hat{\mathcal{H}}_{MF} = 0$) \mgo~lattice~\cite{Cowley88:13}. As predicted by the projection theorem of angular momentum~\cite{Suhonen2007,sakurai:book,Abragam:book}, the energy splitting within the $j\rm{_{eff}}=\frac{1}{2}$ manifold calculated from $\hat{\mathcal{H}}_{pair}$ (Fig.~\ref{fig:MARI}(d)) is linear when $|\lambda| \gg |J|$ with $\widetilde{\alpha}=\frac{50}{9}$~\cite{buyers71:4,Sakurai167:68}.  
%\hh As illustrated in Fig.~[], 
%in the limit of $J \ll \lambda$, the calculated energy splitting within the $j\rm{_{eff}}=\frac{1}{2}$ manifold $\Delta E\left(\left|\frac{J}{\lambda}\right|\right)$ is linear with $\widetilde{\alpha}=\frac{50}{9}$. Such behaviour is consistent with the projection theorem of angular momentum. Corresponding to the Wigner-Eckart theorem for the special case of vector operators ($k=1$), the projection theorem given by 
%\begin{equation}
%\hat{\mathbf{V}} = \frac{\langle\hat{\mathbf{j}}\cdot\hat{\mathbf{V}}\rangle}{j(j+1)}\hat{\mathbf{j}}
%\end{equation}  
%\noindent gives proportionality constants $\frac{5}{3}$ and $-\frac{2}{3}$ for the vector operators $\hat{\mathbf{V}}$ of $\hat{\mathbf{S}}$ and $\hat{\mathbf{l}}$, respectively, for the ground state doublet $j \equiv j\rm{_{eff}} = \frac{1}{2}$ of Co$^{2+}$ ($s=\frac{3}{2}$ and $l = 1$). The proportionality constant of $\frac{5}{3}$ when $\hat{\mathbf{V}} =\hat{\mathbf{S}}$ corresponds to the $\beta$ projection factor mentioned above, thus the projection theorem predicts an effective projection factor $\tilde{\alpha}=2\beta^{2} = 2\left(\frac{5}{3}\right)^{2} = \frac{50}{9}$, in full agreement with the numerically calculated result in the $J \ll \lambda$ limit. 
Therefore, pair excitations in dilute \mgo~provides a direct experimental route for determining $|J|$. The limitation to a magnitude is based on principle that $\Delta E$ is simply an energy difference between the two eigenvalues of the ground state doublet, corresponding to either a transition from $\Gamma\rm{_{eff}}=1$ to $\Gamma\rm{_{eff}}=0$ state or \emph{vice versa}, and thus its value is independent of the sign of $J$~\cite{kittel_thermal:book}. As will be discussed later, the determination of the sign of $J$ is more subtle, ultimately relying on statistical mechanical arguments. \\
\hh While the excitation energy provides $|J|$, the $Q$ dependence can be used to extract the corresponding intra-pair distance $R_{m}$, where $m$ denotes the coordination shell.  By applying the powder-averaged first moment sum rule and the single mode approximation, the $Q$ dependence of the energy-integrated dynamic structure factor for an isolated Co$^{2+}$ pair is proportional to $|F(Q)|^{2}\left(1 - \frac{\sin(QR_{m})}{QR_{m}} \right)$~\cite{hohenberg74:10,stone02:65,stone01:64,plumb16:6,xu00:84,stock09:103,wallington15:92,ma92:69}. Since the modulation is solely dependent on $R_{m}$, an excitation can be assigned to a particular pair and corresponding coordination shell~\cite{Zaliznyak:book,haraldsen05:71}.
%\begin{equation}
%S(Q) \propto \frac{|F(Q)|^{2}}{\Delta E_{o}}\left(1 - \frac{\sin(QR_{m})}{QR_{m}} \right),
%\label{eq:hohenberg} 
%\end{equation}
%\noindent where $F(\mathbf{Q})$ has been approximated here by the isotropic magnetic form factor $F(Q)$.  Since the modulation is solely dependent on $R_{m}$, an excitation can be assigned to a particular pair and corresponding coordination shell.

\subsection{Experimental determination of $J$}

\begin{figure}[htb!]
	\centering
	\scalebox{0.4}{\includegraphics{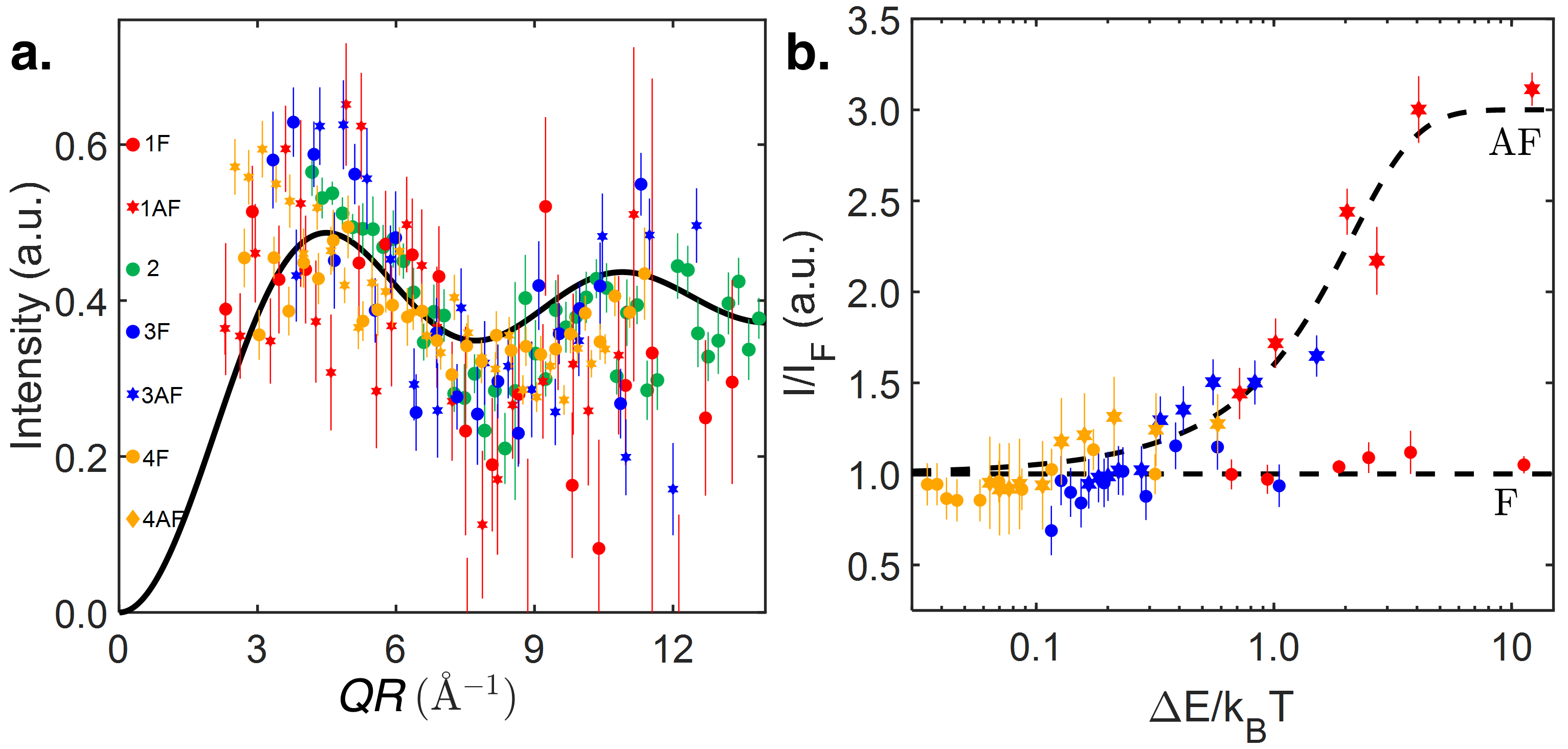}}
	\caption{(a) Form-factor-corrected neutron scattering intensity $Q$ dependence for all seven magnetic excitations previously identified in Fig.~\ref{fig:MARI}. The $Q$ dependence for each magnetic excitation has been rescaled by the intra-dimer distance $R$ obtained from the first moment sum rule~\cite{hohenberg74:10,stone02:65,stone01:64,plumb16:6,xu00:84,stock09:103,wallington15:92,ma92:69}. The black curve corresponds to $1-\frac{\sin(QR)}{QR}$. (b) Normalised emperature dependence of the integrated intensity for the low energy magnetic excitations. All seven excitations fall onto one of two universal curves calculated for antiferromagnetic and ferromagnetic coupled pairs. All quantities were normalised by the analytical expression for the temperature dependence of ferromagnetically coupled pairs, $I_{F}(T )$~\cite{stone08:100,kittel_thermal:book,Zaliznyak:book,xu00:84}. Adapted from Reference~\cite{Sarte98:18} with permission.}  
	\label{fig:dimers}
\end{figure}

\hh Having established the theoretical framework to describe the excitations from the Co$^{2+}$ pairwise interactions in \mgo, the extraction of the values for $J$ from the low energy hierarchy that was measured~\cite{Sarte98:18} on MARI and IRIS will now be addressed. As illustrated in Fig.~\ref{fig:MARI}(d), exchange constants were assigned to each of seven excitations based on the value of excitations' measured energy transfers \emph{via} the diagonalisation of $\hat{\mathcal{H}}_{pair}$ (Eq.~\ref{eq:pair}) using the extracted value of $\lambda=-16(3)$~meV~\cite{Cowley88:13}. By fitting the energy-integrated intensity of the excitations to the first moment sum rule~\cite{hohenberg74:10} (Fig.~\ref{fig:dimers}(a)), each of the 7 exchange constants was assigned to one of the relative coordination shells ranging from $m$ = 1 to $m$ = 4. \\
\hh Inspired by neutron scattering studies on the dimer compound ${\mathrm{Ba}}_{3}{\mathrm{Mn}}_{2}{\mathrm{O}}_{8}$~\cite{stone08:100,stone08:77,Stone11:23}, an experimental route based on the temperature dependence of the integrated intensity was used to determine the sign of $J$. Since the integrated intensity scales as the thermal population difference between the ground and excited states~\cite{stone08:100,kittel_thermal:book,Zaliznyak:book,xu00:84}, the temperature dependence of the integrated intensity for antiferromagnetically coupled ($J>$~0) pairs and its ferromagnetic counterpart are expected to exhibit distinctly different dependence on temperature. As illustrated in Fig.~\ref{fig:dimers}(b),  it was shown that all integrated intensities fall onto either one of two universal curves describing antiferromagnetism or ferromagnetism. The extracted values of $J$ based on the energy, momentum, and temperature dependence are summarised in Tab~\ref{tab:0}. 

\subsection{The ``dual'' hierarchy}

\begin{figure}[htb!]
	\centering
	\scalebox{0.475}{\includegraphics{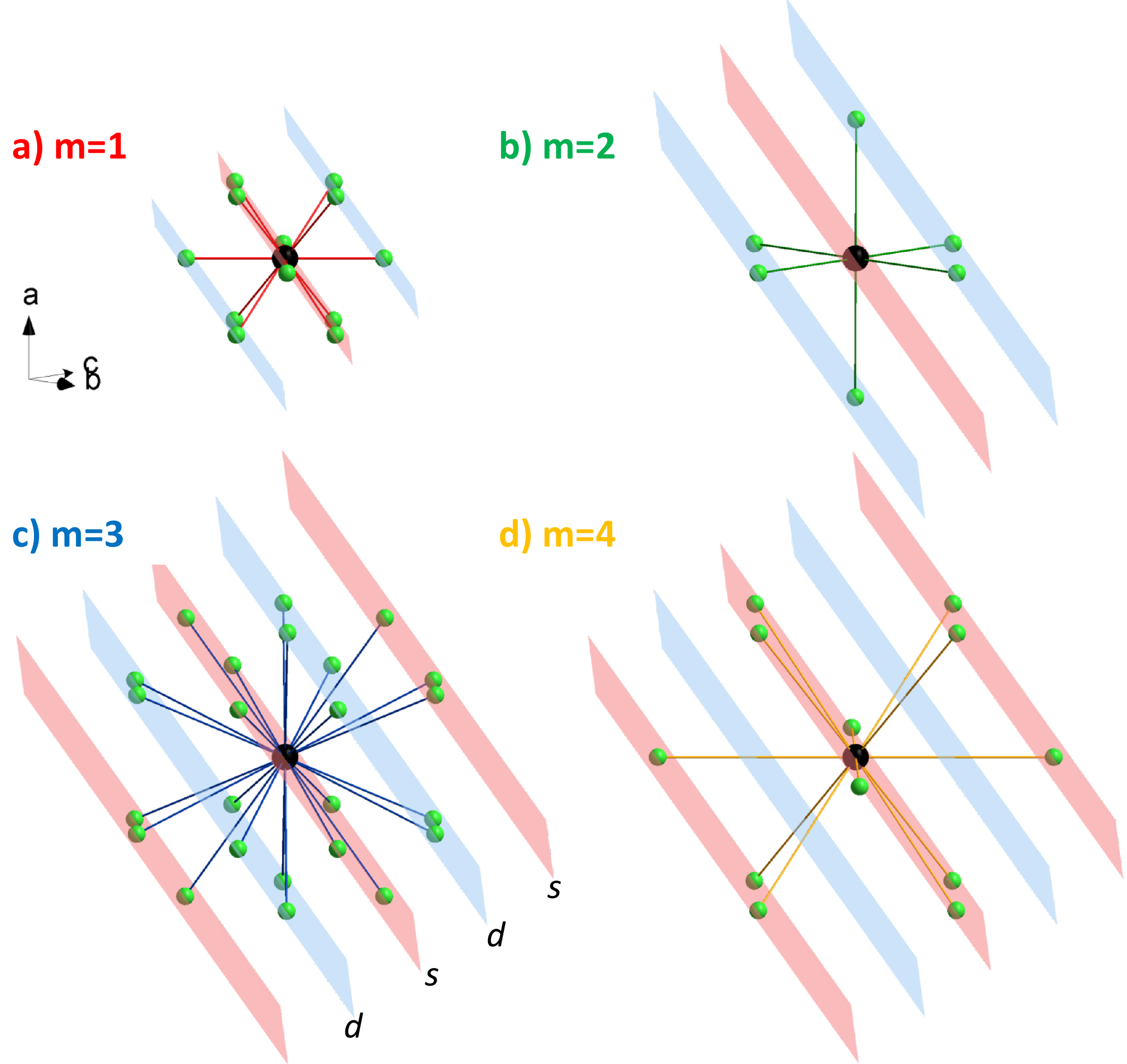}}
	\caption{Isometric view of all Co$^{2+}$ cations located in the (a) first ($m=1$), (b) second ($m=2$), (c) third ($m=3$), and (d) fourth ($m=4$) coordination shells of
		the CoO rocksalt structure (Fig.~\ref{fig:rocksalt}). For the purposes of reference, all (111) planes are labelled as either $s$ and $d$ planes with respect to the reference Co$^{2+}$ (central black site). Adapted from Reference~\cite{Sarte19:100} with permission.}  
	\label{fig:jacky_figure}
\end{figure}

\hh Based on the literature available at the time~\cite{satija01:21,oles06:96}, the observation that all coordination shells, with the exception of $m$ = 2, display two closely spaced excitations with differing signs for $J$ was initially confounding. Although unexpected, the presence of ``dual'' ferro- and antiferromagnetic interactions is in fact consistent with the GKA rules~\cite{goodenough55:100,goodenough58:6,anderson50:79,kanamori59:10,weihe97:36}, since each of these exchange pathways consists of at least one 90$\rm{^{\circ}}$ Co$^{2+}$-Co$^{2+}$ exchange interaction involving the overlap of half and filled $t_{2g}$ orbitals. As illustrated in Fig.~\ref{fig:MARI}(d,inset), the GKA rules indeed predict that the combination of the orbital degree of freedom for each Co$^{2+}$ and a lack of orbital ordering (or anisotropy) in \mgo~would manifest itself as either a direct antiferromagnetic $t^{1}_{2g}$-$t^{1}_{2g}$ or a weaker ferromagnetic $t^{1}_{2g}$-$t^{2}_{2g}$ direct exchange interaction. \\ 
\hh As is summarised in Tab.~\ref{tab:0}, the experimental values of $J$ are not only consistent with the GKA rules, as the antiferromagnetic interaction is stronger than its ferromagnetic counterpart for all the $m\neq2$ excitations, whilst the 180$\rm{^{\circ}}$ Co$^{2+}$-O$^{2-}$-Co$^{2+}$ $m=2$ coupling leads only to a strong antiferromagnetic interaction, but futhermore,the values of $J$ show excellent agreement with general trends reported by both experiment~\cite{tomiyasu06:75,feygenson11:83,daniel69:177,chou76:13,El_Batanouny_2002,satoh17:8,keshavarz18:97,rechtin72:5} and recent GGA+$U$ DFT calculations~\cite{deng10:96}. \\
\hh Finally, the mean field definition of the Curie-Weiss temperature given by~\cite{kittel:book,lee14:35,Sarte98:18_2}
\begin{equation}
\theta{\rm{_{CW}}} = -\frac{2}{3}s(s+1)\sum\limits_{i}z_{i}J_{i},
\label{eq:CW} 
\end{equation}
\noindent enables a direct comparison of the extracted values of $J$ to thermodynamic data for both \mgo~and CoO. In the case of the former, the concurrent presence of AFM and FM implies that the value of $J_{2}$ will ultimately determine $\theta\rm{_{CW}}$. Therefore, by inserting the values of $s=\frac{3}{2}$, $J_{2}= 35.9(6)$~K (3.09(5)~meV), setting the value of  $z_{2}$ to 1 since the chemical dilution reduces the cooperative magnetism to individual
pairwise interactions, and dividing by a correction factor of $\sim1.9$ that was deduced by Kanamori~\cite{kanamori17:57} to account for spin-orbit coupling, Eq.~\ref{eq:CW} yields a value of $-44.9(7)$~K, in agreement with the experimentally determined value of $-41(6)$ K~\cite{Sarte98:18}. \\
\hh In the case of CoO, such a comparison is slightly more involved, and involves the key assumption that CoO assumes a type-II collinear antiferromagnetic structure, corresponding to (111) ferromagnetic sheets that are stacked antiferromagnetically along [111]~\cite{vanLaar65:138,vanLaar66:141,Roth58:110,saito66:21,deng10:96,Herrmann78:11}. As illustrated in Fig.~\ref{fig:jacky_figure}, the determination of $z_{i}$ was accomplished by assigning F and AF labels to Co$^{2+}$ cations that are located on an even integer (or same) number and an odd number of (111) planes away from a reference Co$^{2+}$, respectively. By inserting the values of $s=\frac{3}{2}$, $J_{i}$ as determined from the measured energy transfers, $z_{i}$ from the procedure outlined above, and Kanamori's second perturbation theory correction factor into Eq.~\ref{eq:CW}, a Curie-Weiss temperature of $-295(5)$~K is obtained, in good agreement with both the experimental value of $\theta\rm{_{CW}}=-330$~K~\cite{singer56:104,nagamiya55:4,tachiki64:19,Blanchetais51:12}, and in particular, $T_{N}$ = 291~K~\cite{jauch01:64,Sakurai167:68,kanamori17:57,kanamori17:57_2,trombe51:12,Khan68:29,Salomon74:35,kleinclauss81:14}.  \\
\hh As will be discussed in the next section, the determination of $J$ represents the parameterisation of the last key contribution to Roger's minimalist Hamiltonian (Eq.~\ref{eq:hamiltonian}), laying the foundation for the model that would ultimately be used to address the low energy magnetic fluctuations of CoO.    

\section{Spin-Orbit Excitons in CoO}
\hh  In this section, we describe how Roger's collaborators Sarte \emph{et al.}~\cite{Sarte19:100} extended previous theoretical work on PrTl$_{3}$~\cite{buyers75:11} to construct a mean-field and multilevel spin-orbit exciton model based on Green's functions to address the complex magnetic excitation spectrum of CoO. By employing the spin-orbit and spin exchange  coupling parameters that were previously determined experimentally~\cite{Cowley88:13,Sarte98:18}, the model, based on a tetragonally distorted type-II antiferromagnetic unit cell, successfully captured both the sharp low-energy excitations at the magnetic zone centre, and the energy broadened peaks at the zone boundary. Despite the model's success at low energy transfers, the failure of the model to describe the higher energy excitations leaves an important avenue open for future investigations. 

\subsection{Mean-field theory for multi-level spin-orbit excitons}

\hh The theoretical framework used to describe the localised magnetic response in CoO is based on the Fourier transformed equation-of-motion 
\begin{equation}
\omega G(\hat{A},\hat{B},\omega) = \langle [\hat{A},\hat{B}] \rangle + G([\hat{A},\hat{\mathcal{H}}],\hat{B},\omega),
\label{eq:10}
\end{equation}
\noindent for the response function $G^{\alpha \beta}$ of generic spin operators $\hat{A}$ and $\hat{B}$ that is defined as  
\begin{equation}
G^{\alpha\beta}(i,j, t)  =  -i\Theta(t)\langle[\hat{A}^{\alpha}(i,t),\hat{B}^{\beta}(j,0)]\rangle,
\end{equation}
\noindent and proportional to the magnetic neutron cross section by the fluctuation-dissipation theorem~\cite{tyablikov59:11,Sarry80:23,zubarev60:71,kumar83:28,Shirane:book,Zaliznyak:book}. It is the presence of $[\hat{A},\hat{\mathcal{H}}]$ in Eq.~\ref{eq:10} that necessitates a thorough understanding of $\hat{\mathcal{H}}$, its commutator with $\hat{A}$, and why its parameterisation has dominated the conversation so far.  \\
\hh As first outlined by Buyers~\emph{et al.}~\cite{buyers75:11} and described later in more detail by Sarte~\emph{et al.}~\cite{Sarte19:100}, the derivation of an expression for the neutron response function begins by first rotating all components of the spin operators comprising the inter-ion Hamiltonian $\hat{\mathcal{H}}_{2}$ onto a basis consisting of the eigenstates of $\hat{\mathcal{H}}_{1}$. It can be shown that such a rotation reduces Eq.~\ref{eq:10} to four commutators, and when combined with the \emph{random phase decoupling method}~\cite{wolff60:120,cooke73:7,yamada67:22,yamada66:21} , allows the equation-of-motion to be written as   
\begin{equation}
G^{\alpha\beta}(\mathbf{Q},\omega) = g^{\alpha\beta}(\omega) + g^{\alpha+}(\omega)J(\mathbf{Q})G^{-\beta} (\mathbf{Q},\omega) +  g^{\alpha-}(\omega)J(\mathbf{Q})G^{+\beta} (\mathbf{Q},\omega) + 2g^{\alpha z}(\omega)J(\mathbf{Q})G^{z\beta} (\mathbf{Q},\omega),
\label{eq:alphabeta} 
\end{equation}
\noindent describing the coupling of the single-site response function $g^{\alpha\beta}(\omega)$ by the Fourier transform of the exchange interaction $J(\mathbf{Q})$~\cite{turek06:86}. With the only nonzero single-site response functions in such a highly symmetric environment being: $g^{+-}$, $g^{-+}$, and $g^{zz}$~\cite{holden74:9,buyers71:4,holden74:10}, the restriction of $\alpha\beta$ combinations to $+-$, $-+$, or $zz$ allows Eq.~\ref{eq:alphabeta} to be rewritten as 
\begin{equation}
G^{\alpha\beta}_{ij}(\mathbf{Q},E) = \delta_{ij}g^{\alpha\beta}_{i}(E) + \sum\limits_{k}g^{\alpha\beta}_{i}(E)\Phi J_{i,i+k}(\mathbf{Q})G^{\alpha\beta}_{i+k,j}(\mathbf{Q},E),  
\end{equation} 
\noindent where the prefactor $\Phi=1$ when $\alpha$ = $+$,$-$  or 2 when $\alpha$ = $z$. By approximating CoO as a type-II collinear antiferromagnet~\cite{vanLaar65:138,vanLaar66:141,Roth58:110,saito66:21,deng10:96,Herrmann78:11}., the site indices $i$ and $j$ are restricted to values 1 or 2, yielding four coupled equations for each $\alpha\beta$ combination. These equations when combined together yield  
\begin{equation}
\begin{split}
G^{+-}(\mathbf{Q},E) \equiv \sum\limits_{ij}G^{+-}_{ij}(\mathbf{Q},E)  = 
{{g^{+-}_{1}(E)+g^{+-}_{2}(E)+2g^{+-}_{1}(E)g^{+-}_{2}(E)[J_{d}(\mathbf{Q}) -J_{s}(\mathbf{Q})]} \over {[1-g^{+-}_{1}(E)J_{s}(\mathbf{Q})]\cdot [1  -g^{+-}_{2}(E)J_{s}(\mathbf{Q})]-g^{+-}_{1}(E)g^{+-}_{2}(E)[J_{d}(\mathbf{Q})]^{2}}},\\ 
G^{zz}(\mathbf{Q},E) \equiv \sum\limits_{ij}G^{zz}_{ij}(\mathbf{Q},E)  = 
{{g^{zz}_{1}(E)+g^{zz}_{2}(E)+4g^{zz}_{1}(E)g^{zz}_{2}(E)[J_{d}(\mathbf{Q}) -J_{s}(\mathbf{Q})]} \over {[1-2g^{zz}_{1}(E)J_{s}(\mathbf{Q})]\cdot [1 -2g^{zz}_{2}(E)J_{s}(\mathbf{Q})]-4g^{zz}_{1}(E)g^{zz}_{2}(E)[J_{d}(\mathbf{Q})]^{2}}},
\end{split}
\label{eq:final}
\end{equation}
\noindent where $G^{-+}(\mathbf{Q},E)$ has the same form as $G^{+-}(\mathbf{Q},E)$ with indices $+$ $\longleftrightarrow$ $-$. Here, $J_{s}$ and $J_{d}$ denotes $J(\mathbf{Q})$ on the same ($i=j$) and different ($i\neq j$) sublattices (Fig.~\ref{fig:jacky_figure}), respectively, while $\omega$ has been relabelled as $E=\hbar\omega$, since $\hbar$ is conventionally set to 1. \\
\hh Finally, by the fluctuation-dissipation theorem, the imaginary part of the total response function $G(\mathbf{Q},E) = G^{+-}(\mathbf{Q},E)+G^{-+}(\mathbf{Q},E)+G^{zz}(\mathbf{Q},E)$ is proportional to the dynamical structure factor~\cite{czachor01:63,Vasko05:book,ODASHIMA17:39,Economou:book}, thus providing a direct comparison between the calculated model and experiment.

%\noindent where $\omega$ has been relabelled as $E=\hbar\omega$, and $\hbar$ has been set to 1. Here, the $T\rightarrow0$~K form of the single-site response function has been employed since the energy transfers under consideration $\Delta E \gg T\rm{_{sample}}$. The denominator of $g^{\alpha\beta}$ consists of $E_{no}=\hbar \omega_{n}-\hbar \omega_{0}$ corresponding to the energy associated with the $|0\rangle\rightarrow|n\rangle$ transition, while the presence of the positive infinitesimal $\Delta$ is to ensure analyticality. Coupling between $g^{\alpha\beta}$, and thus the dispersion of $G^{\alpha\beta}$ is defined by $J(\mathbf{Q})$ which is parameterised by both the exchange constant $J_{kl}$ and displacement vector $\mathbf{d}_{kl}$ between sites $k$ and $l$.  As a first approximation, all calculations have considered the simplest case where the exchange interaction is spatially isotropic, although in general this is not case, owing to the anisotropy of the orbital configuration of Co$^{2+}$. \\

\subsection{Model} 
\hh As presented in Eq.~\ref{eq:final}, $G(\mathbf{Q},E)$ is a function of both $g^{\alpha\beta}(E)$ and $J(\mathbf{Q})$. Since $g^{\alpha\beta}$ is itself a function of $\hat{\mathcal{H}}_{1}$ (Eq.~\ref{single_ion}), the single-site response in the limit of $T\rightarrow0$~K is defined by three parameters: $\lambda$, $\Gamma$, and $H_{MF}$. An initial estimate for $\lambda$ was taken to be its reported value of $-16$~meV~\cite{Cowley88:13}.  An initial estimate for $H_{MF}$ was determined from the experimentally determined $\theta\rm{_{CW}}$~\cite{singer56:104,nagamiya55:4,tachiki64:19,Blanchetais51:12} \emph{via} its mean field definition (Eq.~\ref{eq:CW}), yielding an initial estimate of 64.8~meV. The initial estimate of the distortion parameter $\Gamma$=$-8.76$~meV was determined by scaling the reported value of $\Gamma$ for KCoF$_{3}$~\cite{buyers71:4} by the ratio of their respective tetragonal distortions. \\
\hh In the case of $J(\mathbf{Q})= \sum\limits_{i \neq j}J_{ij}e^{i\mathbf{Q}\cdot\mathbf{d}_{ij}}$, it is both a function of $J$ and the magnetic structure under consideration~\cite{turek06:86,tung11:83}. As a first approximation, CoO was treated as a tetragonally distorted type-II collinear antiferromagnet with isotropic exchange constants equal to those reported for Mg$_{0.97}$Co$_{0.03}$O~\cite{Sarte98:18}. In order to account the possibility for both the ``dual hierarchy'' and the tetragonal distortion, 16 different orbital configurations of the form $xAxx\gamma$ were considered. Whereas $x$ could either be antiferromagnetic (A) or ferromagnetic (F) to account for the possibility for either anti-/ferromagnetic coupling in coordination shells $m=$1, 3, and 4,      the index $\gamma$ could assume labels 1 or 2, distinguishing the presence or absence of distorted bonding configurations. Finally, in order to account the large experimental beam present on MERLIN~\cite{BEWLEY06:385}, the model considered the mean contribution from all 16 equally weighted $xxAx\gamma$ orbital configurations, each with the same value of $\lambda$, $\Gamma$, and $J_{n}$ that are subject to different values of $H_{MF}$, corresponding physically to a unique type of ``domain" in the bulk CoO single crystal. 

\subsection{Comparison between model \& experiment: success and failures}

\begin{figure}[htb!]
	\centering
	\scalebox{0.285}{\includegraphics{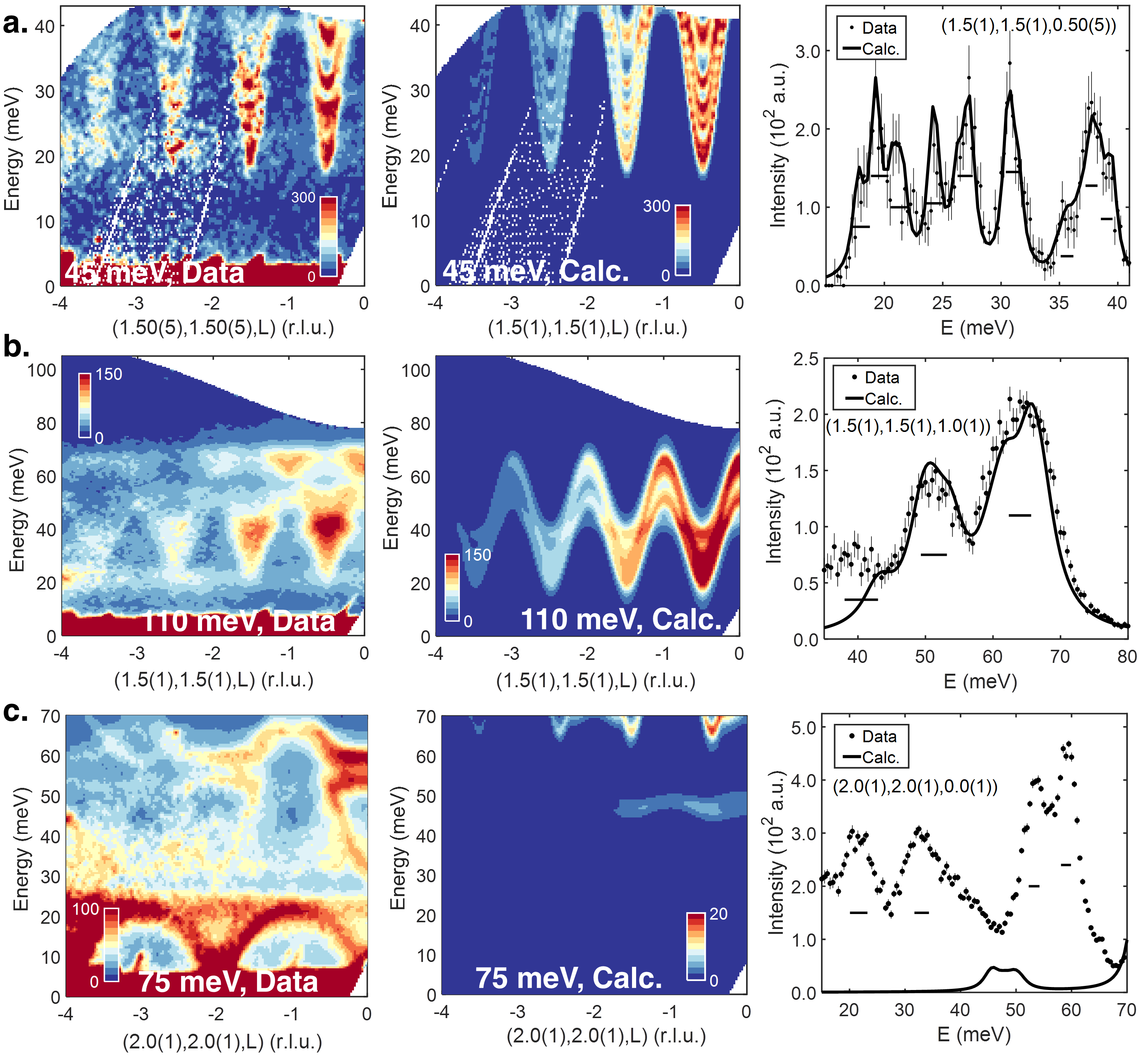}}
	\caption{(a) $(\mathbf{Q},E)$ slices of CoO measured by Sarte \emph{et al.}~\cite{Sarte19:100} on MERLIN at 5~K. (b) Corresponding $(\mathbf{Q},E)$ slices calculated with a mean-field multi-level spin-orbit exciton model~\cite{buyers75:11} employing the refined parameters listed in Tabs.~\ref{tab:1} and~\ref{tab:2}. All $(\mathbf{Q},E)$ slices have been folded along [001].  (c) Comparison between $\mathbf{Q}$-integrated cuts of (a) and (b).  Horizontal bars in $\mathbf{Q}$-integrated cuts indicate instrumental resolution~\cite{pychop}. Adapted from Reference~\cite{Sarte19:100} with permission.}  
	\label{fig:calculations}
\end{figure}

\begin{figure}[htb!]
	\centering
	\scalebox{0.25}{\includegraphics{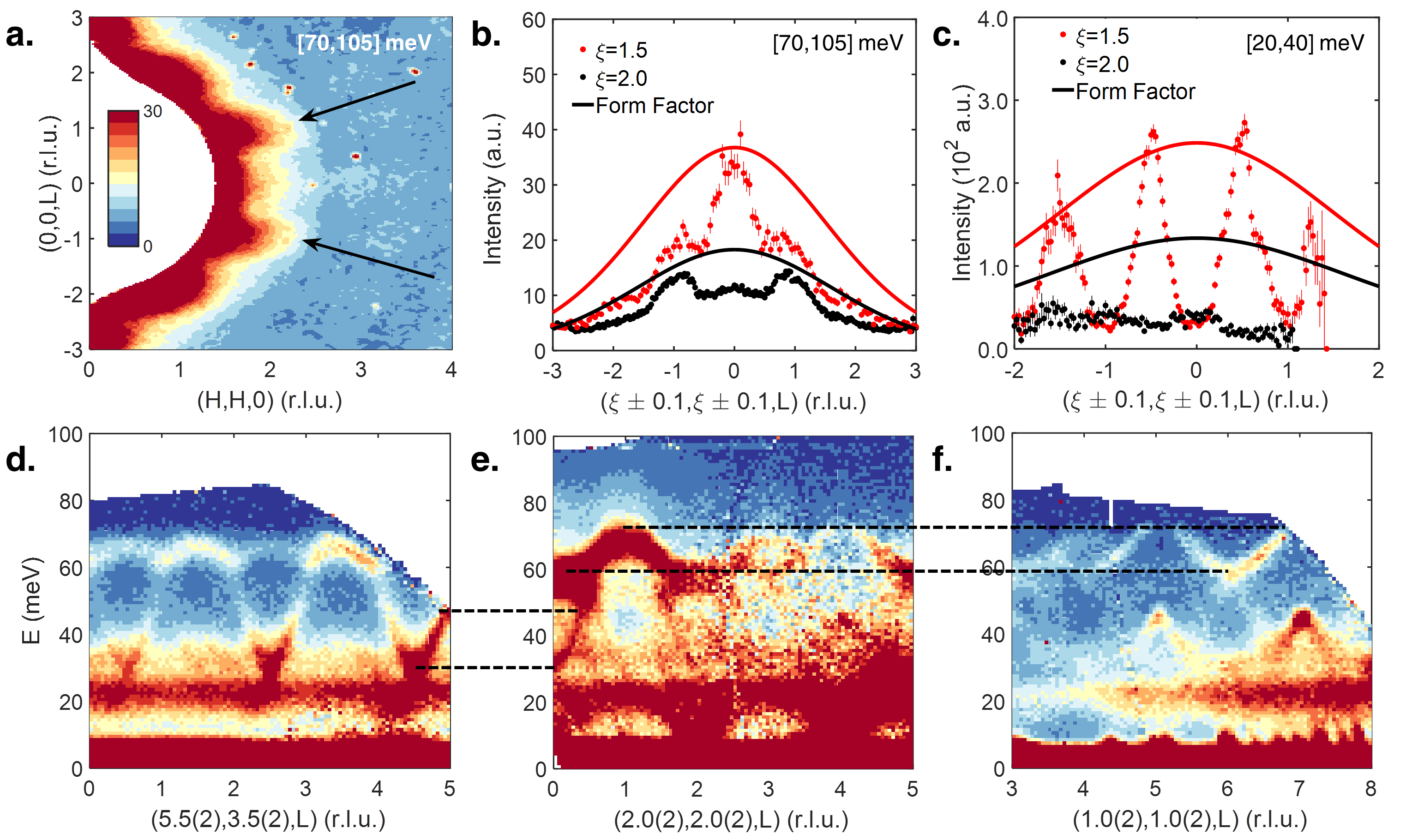}}
	\caption{(a)  Energy-integrated ($E=[70,105]$~meV) slice measured on MERLIN by Sarte \emph{et al.}~\cite{Sarte19:100} at 5~K with an $E_{i}$ of 110~meV.  Arrows indicate fluctuations that may exhibit itinerant-like behaviour, as (b) a $\mathbf{Q}$-integrated ($\xi$,$\xi$,$L$) cut of (a) falls off more rapdily than the Co$^{2+}$ magnetic form factor, in stark contrast with the behaviour of (c) an equivalent $\mathbf{Q}$-integrated ($\xi$,$\xi$,$L$) cut for an $E_{i}=45$~meV.  Comparison between the phonon scattering at large $Q$ centred about (d,f) the nuclear zone boundaries and (e) modes located near the magnetic zone boundary along the $(2,2,0)$ direction that were unaccounted for by the spin-orbit exciton model~\cite{Sarte19:100}. Dashed lines indicate the overlap of the energy transfers range between both gapped optical phonon modes centered near the nuclear zone boundaries and the low-$Q$ scattering near the $(2,2,0)$ magnetic zone boundary. All three $(\mathbf{Q},E)$ slices have been folded along [001] and have been renormalised to share a common relative intensity scale. Adapted from Reference~\cite{Sarte19:100} with permission.}  
	\label{fig:failures}
\end{figure}

\hh As illustrated in Fig.~\ref{fig:calculations} and summarised in Tabs.~\ref{tab:1} and~\ref{tab:2}, by allowing the value of $H_{MF}$ to refine independently for each of the equally weighted 16 $xAxx\gamma$ domains, with each possessing identical refined values of $\lambda$, $J$, and $\Gamma$, the spin-orbit exciton model~\cite{Sarte19:100,buyers75:11} successfully reproduced both the fine structure at the magnetic zone centre and the broad excitations at the $(1.5,1.5,-1)$ zone boundary, whilst capturing the steeply dispersive columns of scattering observed at higher energy transfers. \\
\hh The need for all 16 $xAxx\gamma$ domains stems from the low energy fine structure at the zone centre~\cite{yamani403:08,yamani88:10,tomiyasu06:75}. Although an individual $xAxx\gamma$ domain does reproduce the correct bandwidth, the neutron response is dominated by a single highly dispersive $G^{-+}$ mode, while both $G^{-+}$ and $G^{zz}$ modes are significantly weaker in intensity and weakly dispersive at higher energy transfers. By considering the fine structure to consist of multiple overlapping $G^{-+}$ modes exclusively, other additional features of the data are captured by the model including: the flat band at $\sim$40~meV, the weak dispersionless mass of scattering at $\sim$80-90~meV, and the steeply dispersive columns of high energy scattering at the $(1.5,1.5,-1)$ zone boundary. \\
\hh In addition to the necessity for 16 $xAxx\gamma$ domains, the discrepancy between the calculated dispersion and that measured experimentally indicated the need for the optimisation of the model's parameters. Corresponding to the main parameter for $\hat{\mathcal{H}}_{pair}$, $\lambda$ plays a key role in the determination of $|J|$, in particular $|J_{2}|$, from the experimentally determined $\Delta E$. In the case of a fixed value of $H_{MF}$, an increase in the value of $|\lambda|$ corresponds to a decrease in $|J|$ for a given $\Delta E$. The result is a weaker dispersion that shifts the dominant $G^{-+}$ component to higher energy transfers at the zone centre, while minimally affecting the two other weakly dispersive modes. Such behaviour contrasts the effect of $H_{MF}$ which is not on the dispersion itself, but rather a change in its value results in a systematic shift in the value of energy transfers for a fixed $\mathbf{Q}$, with the shift being significantly larger compared for the same relative change for $|\lambda|$. Finally, in contrast with $\lambda$ ($J$) and $H_{MF}$, the influence of the $\Gamma$ is most pronounced $G^{+-}$, providing a mechanism to shift the $G^{+-}$ mode without inducing a comparable shift for the prominent $G^{-+}$ mode when all other parameters are fixed. \\
\hh With such a large number of domains under consideration, constraints on the parameter space for $J$ \emph{via} $\lambda$, $\Gamma$, and $H_{MF}$ were required to ensure convergence for a least squares optimisation. The parameter spaces for both $\lambda$ and $\Gamma$ were defined by the values previously reported in the literature~\cite{Cowley88:13,Abragam:book,tomiyasu06:75,tomiyasu04:70,kant08:78,vanschooneveld12:116,satoh17:8,Austin70:33,buyers71:4,gladney66:146,ferguson63:39,hirakawa60:15,thornley65:284}. Initial modelling attempts fixed the values of $J$ to those obtained from the diagonalisation of $\hat{\mathcal{H}}_{pair}$ for a fixed $\lambda$~\cite{Sarte98:18}. The failure of these initial attempts to reproduce both the fine structure at the zone centre and the broad excitations at the $(1.5,1.5,-1)$ zone boundary simultaneously was ultimately remedied by allowing the values of $J$ to vary $\pm$20\% from its value obtained from $\hat{\mathcal{H}}_{pair}$. In contrast to all the other parameters under consideration that were set to be equal for all 16 domains,  no such restriction was applied to $H_{MF}$ which was allowed to vary from 0 to an arbitrarily large upper limit. \\
\hh Despite the success of the spin-orbit exciton model~\cite{Sarte19:100,buyers75:11} to account for some of the experimental data, its success appeared to be limited to low energy transfers along $(1.5,1.5,L)$. Along $(2,2,L)$, both the spin-orbit exciton model and linear spin wave theory predicted nearly zero intensity for magnetic fluctuations for an antiferromagnetic structure, in stark contrast with the prominent magnetic scattering that was observed experimentally (Fig.~\ref{fig:calculations})~\cite{Sakurai167:68,Austin70:3}. On the other hand, although the model did successfully reproduce the steep columns of scattering at higher energy transfers at the zone boundaries along $(1.5,1.5,L)$, a closer inspection of the $\mathbf{Q}$ dependence of these high energy modes (Fig.~\ref{fig:failures}(a) revealed that their intensity decayed more rapidly than what is predicted by the magnetic form factor $f(Q)$, indicating the presence of delocalised magnetism~\cite{plumb18:97,stock15:114}. While the similarity between the dispersion of phonons at high $\mathbf{Q}$ (Fig~\ref{fig:failures}(b)) and the magnetic scattering absent in the spin-orbit exciton model suggests that a magneto-vibrational contribution~\cite{Egelstaff:book,fennell14:112,Hennion02:312} may account for the model's failures along $(2, 2, L)$, there is no clear physical mechanism to account for the failure at high energy transfers. Promising possibilities include spatial extended magnetism due to strong covalent bonding and significant hybridisation of the $3d$ orbitals that was found to be the case for Sr$_{2}$CuO$_{3}$~\cite{walters09:5}, or possibly multi-magnon decay processes, supported by distinct similarities of the high energy response in CoO to those for itinerant magnets~\cite{ishikawa77:16,lorenzo94:72,fincher79:43,endoh06:75,bruke83:51,sinha77:15,sternlieb93:48,stock14:90,stock07:75,stock10:82,plumb14:89}.

\section{Concluding Remarks} 

\hh The story of CoO is of one that has come full circle. Here we have described how Roger, through his extraction of the single-ion parameters of Co$^{2+}$ \emph{via} chemical dilution~\cite{Cowley88:13,Sarte98:18}, and Bill, through the establishment of the spin-orbit exciton model~\cite{buyers75:11}, laid for the foundation for the future success of Sarte \emph{et al.}~\cite{Sarte19:100} in parameterising the low energy magnetic fluctuations in CoO that were first measured with neutrons by these two at Chalk River more than 50 years ago~\cite{Sakurai167:68}. While the failure of the model at the zone boundaries can be attributed to a magneto-vibrational contribution to the neutron cross section~\cite{Egelstaff:book,fennell14:112,Hennion02:312}, the physical origin underlying the rapid decay of the column of magnetic fluctuations at high energy transfers remains an open question. Despite the strong insulating nature of CoO~\cite{vanElp44:91}, it can be speculated that the model's failures at high energy transfers corresponds to the breakdown of spin-orbit excitons which may be accompanied by a crossover from localised to spatially-extended magnetism, reminisicent of an itinerant-like response~\cite{stock15:114} or strong covalency~\cite{walters09:5}.    \\
\hh The success of the spin-orbit exciton model~\cite{buyers75:11,Sarte19:100} with such a simple Hamiltonian in addressing the low energy magnetic fluctuations for a system with an orbital degree of freedom that is far displaced from the $\lambda \gg J$ limit such as CoO, suggests that such a model would be of great interest for a community who has recently paid particular interest on Mott insulators with strong spin-orbit coupling in the search for unconventional, and often novel magnetic ground states~\cite{hogan16:93,jackeli09:102,Kim09:1323,khaliullin04:93,katsura05:95,haverkort08:101,liu08:101}. The parameters extracted from this analysis will be of importance in future work in understanding the response of exchange bias in thin films and also in resolving the low temperature nuclear and magnetic structures in CoO.  Representing a powerful tool that allows for the direct parameterisation of a simple Hamiltonian to model the excitation spectrum under consideration, it would be of interest to see if the spin-orbit exciton model~\cite{buyers75:11} would be able to achieve the same success in addressing low energy magnetic fluctuations in high $Z$ magnets ($e.g.$ Ru$^{4+}$, Ir$^{4+}$) as calculations currently reported in the literature that are based on linear spin wave theory, often employing very complex Hamiltonians~\cite{jain17:13,kim12:109,wang13:3,RAPP01:324,buyers06:62,coldea01:86,roger89:39,macdonald88:37,wysin15:book,datta12:85,xiao13:87,khomskii:book,Arts:book,yosida:book}.

\newpage

\ack{
\hh We would like to convey our sincerest gratitude to the late Roger~Cowley. His unrelenting pursuit of a minimalist approach, rivalled only by his unwaivering productivity, laid the foundation for future progress in addressing the low energy magnetic excitations, with much of the progress that would be seen in the years following Roger's passing being facilitated by the assistance and guidance from his long term collaborator and close friend Bill~Buyers. \\
\hh From sample preparation to data aquisition to data analysis, we would like to recognise the key roles played by our many collaborators throughout the years. We convey our many thanks to:  D.~Le, V.~Garc\'{i}a-Sakai, R.~A.~Ewings, J.~W.~Taylor, C.~D.~Frost, D.~Prabhakaran, E.~E.~Rodriguez, Z.~Yamani, M.~Songvilay, E.~Pachoud, K.~H.~Hong, A.~J.~Browne, A.~Kitada, C.~MacEwen, G.~M.~McNally, G.~Perversi, E.~J.~Pace, B.~R.~Ortiz, C.~Schwenk, and K.~J.~Camacho. \\
\hh Finally, we would like to acknowledge financial support from the Royal Society, NSF, STFC, ERC, EPSRC, and the Carnegie Trust for the Universities of Scotland.~P.~M.~S. acknowledges
financial support from the University of California, Santa Barbara through the Elings Fellowship, and the University of Edinburgh
through the GRS and PCDS. A portion of this work was supported by the DOE, Office of Science, Basic Energy Sciences under Award DE-SC0017752.} 
\section*{References}

\bibliographystyle{iopart-num}
%\bibliography{references_cowley}
\providecommand{\newblock}{}

\newpage
\renewcommand\thesection{}
\section*{Figures}

\FloatBarrier
\newpage
\section*{Tables}

\begin{table}[htb!]
	\centering
	\caption{Comparison of parameter values experimentally determined from \mgo~by Sarte \emph{et al.}~\cite{Sarte98:18} and their values for CoO as cited in literature~\cite{tomiyasu06:75,feygenson11:83,daniel69:177,chou76:13,El_Batanouny_2002,satoh17:8,keshavarz18:97,rechtin72:5,Smart:book} and calculated by Deng \emph{et al.}\cite{deng10:96} using GGAs$+U$ DFT. The values of $J$ that were calculated by DFT have been renormalised such that $J_{2}$ has been set equal to its experimentally determined value for \mgo. The values of T$\rm{_{N}}$, $\theta\rm{_{CW}}$ and $\lambda$ reported in literature~\cite{jauch01:64,Sakurai167:68,kanamori17:57,kanamori17:57_2,trombe51:12,Khan68:29,Salomon74:35,kleinclauss81:14,singer56:104,nagamiya55:4,tachiki64:19,Blanchetais51:12,Cowley88:13,ferguson63:39,hirakawa60:15,thornley65:284} for CoO have been included for the purposes of a comparison to the mean field value of $\theta\rm{_{CW}}$ that was determined from the experimental values of $J$ for \mgo. Numbers in parentheses indicate statistical errors.}
	\begin{tabular}{|c|c|c|c|}
		\hline
		Parameter	& Sarte \emph{et al.}~\cite{Sarte98:18} (meV)  & Literature (meV)  & Deng \emph{et al.}~\cite{deng10:96} (meV)   \\ 
		\hline
		$\lambda$ & & $-12$ to $-22.1$~\cite{Cowley88:13,Abragam:book,tomiyasu06:75,tomiyasu04:70,kant08:78,vanschooneveld12:116,satoh17:8,Austin70:33,ferguson63:39,hirakawa60:15,thornley65:284} &\\ 
		\hline
		$J_{1AF}$ & $-1.000(8)$ & \multirow{ 2}{*}{$-0.60$ to 0.31~\cite{tomiyasu06:75,feygenson11:83,daniel69:177,chou76:13,El_Batanouny_2002,satoh17:8,keshavarz18:97,rechtin72:5,Smart:book}} & \multirow{ 2}{*}{0.97(2)} \\ \cline{1-2}
		$J_{1F}$ & 0.918(6)  &    & \\  
		\hline
		$J_{2}$ or $J_{2AF}$ & $-3.09(5)$  & $-2.8$ to $-0.0013$~\cite{tomiyasu06:75,feygenson11:83,daniel69:177,chou76:13,El_Batanouny_2002,satoh17:8,keshavarz18:97,rechtin72:5,Smart:book}   & $-3.09(5)$ \\ \hline
		$J_{3AF}$ & $-0.258(1)$ & \multirow{ 2}{*}{0.67~\cite{tomiyasu06:75}}   & \multirow{ 2}{*}{0.461(8)} \\ \cline{1-2}
		$J_{3F}$ & 0.182(1) &  & \\  \hline
		$J_{4AF}$ & $-0.0759(4)$ &   & \multirow{ 2}{*}{0.0085(1)} \\ \cline{1-2}
		$J_{4F}$ & 0.0504(4)  &  & \\ \hline
		$T\rm_{_{N}}$	&   & 25.1(4)~\cite{jauch01:64,Sakurai167:68,kanamori17:57,kanamori17:57_2,trombe51:12,Khan68:29,Salomon74:35,kleinclauss81:14} &  \\ \hline 
		$\theta\rm_{_{CW}}$	& $-25.4(4)$  & $-28.4(4)$~\cite{singer56:104,nagamiya55:4,tachiki64:19,Blanchetais51:12} &  \\ 
		\hline
	\end{tabular}
	\label{tab:0}
\end{table}

\begin{table}[htb!]
\centering
\caption{Summary of the initial values, parameter spaces, and refined values for the parameters of the mean-field multi-level spin-orbit exciton model reported by Sarte \emph{et al.}~\cite{Sarte19:100}. All values are reported in meV and numbers in parentheses indicate statistical errors.}
{\renewcommand{\arraystretch}{1}
\begin{tabular}{|c|c|c|c|}
\hline
Parameter&Initial Value&~~Range~~&Refined Value\\ 
\hline
$\lambda$ & $-16$ & [$-19$,$-13$] & $-19.00(1)$ \\ \hline
$\Gamma$ & $-8.76$ & [$-8.76$,$-6.16$] & $-6.16(1)$ \\ \hline
$J_{1F}$ & $-0.918$ & [$-1.134$,$-0.730$]& $-0.780(1)$ \\ \hline
$J_{1AF}$ & 1.000 & [0.798,1.24] & 0.848(1) \\ \hline
$J_{2}$ & 3.09 & [2.29,4.55]& 2.43(1) \\ \hline
$J_{3F}$ & $-0.182$  & [$-0.220$,$-0.145$]& $-0.154(1)$ \\ \hline 
$J_{3AF}$ & 0.262 & [0.209,0.316]& 0.223(1) \\ \hline
$J_{4F}$ & $-0.0504$ & [$-0.0581$,$-0.0402$]& $-0.0428(1)$ \\ \hline
$J_{4AF}$ & 0.0759 & [0.0606,0.0874] & 0.0645(1) \\ \hline
$H_{MF}$ & 64.8 & [0,100] & Tab.~\ref{tab:2} \\ \hline
\end{tabular}}
\label{tab:1}
\end{table}

\begin{table}[htb!]
\centering
\caption{Refined values (in meV) of the mean molecular field parameter $H_{MF}$ for all 16 $xAxx\gamma$ orbital configurations for the mean-field multi-level spin-orbit exciton model reported by Sarte \emph{et al.}~\cite{Sarte19:100}, with each configuration's parameters equal to the refined values listed in Tab.~\ref{tab:2}. Numbers in parentheses indicate statistical errors.}
{\renewcommand{\arraystretch}{1}
\begin{tabular}{|c|c|}
\hline
~~~~Orbital Configuration~~~~&~~~~Refined Value~~~~\\ 
\hline\hline
AAAA1 & 62.4(2) \\ \hline
AAAA2 &46.2(1) \\ \hline\hline
AAAF1 & 55.3(2) \\ \hline 
AAAF2 & 53.9(1) \\ \hline \hline
AAFA1 & 46.2(1)\\ \hline
AAFA2 & 49.9(1)\\ \hline \hline
AAFF1 & 56.2(2)\\ \hline
AAFF2 & 56.0(2)\\ \hline \hline
FAAA1 & 47.3(1)\\ \hline
FAAA2 & 47.9(1)\\ \hline \hline
FAAF1 & 58.9(3)\\ \hline
FAAF2 & 58.8(2)\\ \hline \hline
FAFA1 & 61.6(3) \\ \hline
FAFA2 & 60.9(3)\\ \hline \hline
FAFF1 & 48.9(1)\\ \hline
FAFF2 & 59.5(2)\\ \hline \hline
Average & 54.4(4)\\ \hline 
\end{tabular}}
\label{tab:2}
\end{table}

\end{document}